\documentclass[aps,pre,preprint,showpacs,amsmath,amsfonts,amssymb]{revtex4-1}
\usepackage{amsmath,amsfonts,amssymb,color}
\usepackage{amsthm}
\usepackage{leftidx}
\usepackage{graphicx}
\usepackage{dcolumn}
\usepackage{bm}
\usepackage{epstopdf}
\usepackage{epsfig}

\newtheorem{lemma}{Lemma}

\newcommand{\n}{\noindent}

\begin{document}
\title{Floquet topological semimetal phases of an extended kicked Harper model}
\author{Raditya Weda Bomantara}
\author{Gudapati Naresh Raghava}
\author{Longwen Zhou}
\author{Jiangbin Gong}%
\email{phygj@nus.edu.sg}
\affiliation{%
Department of Physics, National University of Singapore, Singapore 117543
}%
\date{\today}


\begin{abstract}
Recent discoveries on topological characterization of gapless systems have attracted interest in both theoretical studies and experimental realizations. Examples of such gapless topological phases are Weyl semimetals, which exhibit 3D Dirac cones~(Weyl points), and nodal line semimetals, which are characterized by line nodes~(two bands touching along a line). Inspired by our previous discoveries that the kicked Harper model exhibits many fascinating features of Floquet topological phases, in this manuscript we consider a generalization of the model, where two additional periodic system parameters are introduced into the Hamiltonian to serve as artificial dimensions, so as to simulate a $3$D system. We observe that by increasing the hopping strength and the kicking strength of the system, many new band touching points at Floquet quasienergies $0$ and $\pi$ will start to appear. Some of them are Weyl points, while the others form line nodes in the parameter space. By taking open boundary conditions along the physical dimension, edge states analogues to Fermi arcs in static Weyl semimetal systems are observed. Finally, by designing an adiabatic pumping scheme, the chirality of the Weyl points and the $\pi$ Berry phase around line nodes can be manifested.

\end{abstract}
\pacs{03.65.Vf, 05.60.Gg, 05.30.Rt, 73.20.At}


\maketitle

\section{Introduction}
One paradigm in studies of quantum chaos is the kicked Harper model~(KHM).  It was originally introduced in Ref.~\cite{KHM} to model a kicked charge in a magnetic field.  Later a whole class of generalized versions of KHM were introduced, which may also be realized via a periodically kicked charge in a magnetic field \cite{Dana}.  Remarkably, more than two decades ago, the KHM was already used to study topological phase transitions in Floquet states and to understand possible connections between large Floquet band Chern numbers, delocalization of Floquet eigenstates, and the underlying classical chaos~\cite{Chaos, Chaos2}. Refs.~\cite{Chaos, Chaos2} were perhaps the first two studies illuminating on topological phases of matter in periodically driven systems, a topic of wide interest at present due to the finding of Floquet topological insulators~\cite{FTI}. Studies of topological phase transitions in the KHM opened up stimulating questions of interest to different communities, connecting concepts from different areas such as condensed matter physics and nonlinear dynamics.

The relevance of the KHM to studies of topological phases becomes even more evident in recent years. In particular, it was shown that the KHM and a variant of another paradigm of quantum chaos, namely, the double-kicked rotor model~(DKR), have a certain kind of topological equivalence though their spectrum has many distinctively different features~\cite{Hailong}. Furthermore, it was shown that the KHM has drastically different edge states as compared with that associated with the DKR, even though all their bulk Floquet bands have exactly the same Chern numbers~\cite{Derek}.  This result is useful for ongoing investigations of the issue of bulk-edge correspondence in Floquet topological states as well as the issue of weak topological insulators.

In previous studies~\cite{Derek}, we already showed that adding one periodic system parameter to the conventional KHM Hamiltonian can serve the purpose of introducing an artificial dimension, thus enabling us to study the properties of two-dimensional~($2$D) systems from the perspective of one-dimensional~($1$D) systems. It was found that the quasienergy spectrum of this effective $2$D KHM is able to host as many $2$D Dirac cones as possible at both quasienergy $0$ and quasienergy $\pi$ by simply increasing the kicking strength. This suggests that the KHM might be a good platform to study gapless topological phases, such as Weyl semimetal~\cite{Burkov, Ferarc, Hosur2, Ferarc2} and nodal line semimetal~\cite{Burkov, Aji, Mullen, LN2, Youngkuk} in Floquet systems. Motivated by these results regarding the KHM, in this paper we propose to study an extended KHM by introducing two periodic system parameters into the Hamiltonian. With two periodic system parameters available, together with the conserved quasimomentum variable due to the translational invariance in the sole physical~(lattice) dimension, it is possible to obtain Floquet bands that can be effectively associated with three-dimensional~($3$D) systems. This thus creates an interesting opportunity to explore the KHM from a perspective related to $3$D gapless topological phases of matter, which has recently become an interesting topic~\cite{Parsa, Shunji}.

In order to make both periodic system parameters nontrivial, we propose to introduce modulations to both the lattice hopping term and the kicking potential in the KHM. These modulations are characterized by two phase shift parameters playing the role of two periodic system parameters. The model Hamiltonian can thus also be deemed as a kicked version of a recently proposed off-diagonal Harper model~\cite{Sarma1}. Having two periodic system parameters leads to one direct advantage in understanding the implications of various topological phases for the dynamics. Specifically, it is now possible to execute adiabatic loops in the $2$D parameter space spanned by the two phase shift parameters, by directly and slowly tuning the two system parameters. Therefore, even for a single quasimomentum component of the system, it is possible to realize closed adiabatic loops to manifest the topological phases.

The purpose of this paper is to present a variety of interesting and motivating results for the simplest two-band case of the above-mentioned extended KHM. First of all, we observe the emergence of isolated $3$D Dirac cones~(Weyl points) as well as line nodes~(two bands touching along a line) in the Floquet bands, at quasienergies $0$ and $\pi$. This strongly suggests that even $1$D driven systems can be useful to study $3$D Floquet topological phases. In particular, the emergence of Weyl points and line nodes at quasienergy $\pi$ is a finding that is absent in static systems. Secondly, by taking advantage of the simplicity of the two-band case, we also analytically show how it is possible to obtain as many Weyl points and line nodes as possible in our extended KHM. Thirdly, when we take open boundary conditions~(OBC) along the only physical dimension, we observe a line connecting two Weyl points with opposite chiralities, which is analogues to the so-called Fermi arc~\cite{Ferarc} in static Weyl semimetal systems, at both quasienergies $0$ and $\pi$. Fourthly, we also perform dynamics studies and demonstrate how the chirality of the (Floquet) Weyl points and the $\pi$ Berry phase around the (Floquet) line nodes can be manifested in the adiabatic dynamics of the system. By considering an adiabatic loop with an appropriate initial state in the $2$D parameter space enclosing a Weyl point, its chirality can be captured. Furthermore, by adiabatically driving a Floquet eigenstate at a fixed quasimomentum over one adiabatic cycle, the $\pi$ Berry phase can be observed. These results should be of experimental interest, especially for waveguide and optical lattice realizations of $1$D driven systems.

This paper is organized as follows. In Sec.~\ref{II}, we start by describing the Hamiltonian studied in this paper, which is referred to as the off-diagonal modulated Kicked Harper model~(ODKHM), and a possible experimental realization. By focusing on the situation where the Floquet operator associated with this Hamiltonian consists of only two bands, we analytically show the emergence of band touching points in a form of isolated Weyl points or line nodes as the hopping strength and kicking strength are varied. In Sec.~\ref{III}, we present numerically the emergence of localized edge states associated with the Weyl points and line nodes by taking OBC along the physical or artificial dimension of the system. In Sec.~\ref{IV} A, we design an adiabatic pumping scheme surrounding a Weyl point and show how its chirality can be captured by choosing an appropriate initial state. We further show that by using the same adiabatic pumping scheme and initial state, the chirality of Weyl points at quasienergy $0$ and quasienergy $\pi$ are captured in different manners, which may suggest the distinction between the two types of Weyl point. In Sec.~\ref{IV} B, we also show a possible way to capture the $\pi$ Berry phase of the line node by adiabatically pumping a Floquet eigenstate at a fixed value of its quasimomentum. Finally, we summarize our results and present possible future directions in Sec.~\ref{V}.

\section{Emergence of Weyl points and line nodes in the ODKHM}\label{II}

\subsection{Description of the model}
We start by considering the following Hamiltonian,

\begin{equation}
\begin{split}
&H=\sum_{n=1}^{N-1}\left\lbrace\left[J+\lambda \cos(2\pi \beta_1 n+\phi_y)\right]\, |n+1\rangle \langle n |+h.c.\right\rbrace \\
&+\sum_{n=1}^{N-1}\sum_{j} V \cos(2\pi \beta_2 n+\phi_z) \, |n \rangle \langle n | \, \delta(t-jT)\;,
\end{split}
\label{fund}
\end{equation}

\n where $J$ and $\lambda$ are the parameters controlling the hopping strength, $V$ is the kicking strength, $T$ is the kicking period, $n$ is the lattice site index, and $N$ counts the total number of lattice sites. If $\lambda=0$, the above Hamiltonian reduces to the conventional KHM. When $\lambda\neq 0$, however, Eq.~(\ref{fund}) depends on two periodic parameters $\phi_y$ and $\phi_z$, which can take any value in $[-\pi,\pi)$. Physically, $\phi_y$ and $\phi_z$ represent the phase shift in the hopping and kicking strength modulations, which can be independently tuned. In particular, because of the extra parameter $\phi_y$ which modulates the hopping strength, Eq.~(\ref{fund}) shall be referred to as the off-diagonal Kicked Harper model~(ODKHM). Due to the range of values they can take, the two parameters $\phi_y$ and $\phi_z$ can play the role of quasimomenta along perpendicular directions to simulate a $3$D system. By taking $\beta_1=\beta_2=\frac{1}{2}$, Eq.~(\ref{fund}) becomes,

\begin{equation}
\begin{split}
&H=\sum_{n=1}^{N-1}\left\lbrace\left[J+(-1)^n\lambda \cos(\phi_y)\right]\, |n+1\rangle \langle n |+h.c.\right\rbrace \\
&+\sum_{n=1}^{N-1}\sum_{j} (-1)^nV \cos(\phi_z) \, |n \rangle \langle n | \, \delta(t-jT)\;.
\end{split}
\label{fund2}
\end{equation}

Since Eq.~(\ref{fund2}) is time periodic, we can define a Floquet operator as a one period time-evolution operator from $t=0^{-}$ to $t=T^{-}$, which is given by,

\begin{equation}
U=e^{-\mathrm{i} \sum_{n=1}^{N-1}\left\lbrace\left[J+(-1)^n \lambda \cos(\phi_y)\right] \, |n+1\rangle \langle n |+h.c.\right\rbrace} e^{-\mathrm{i} \sum_{n=1}^{N-1} (-1)^n V \cos(\phi_z)\, |n \rangle \langle n |}\;,
\label{tri}
\end{equation}

\n where we have assumed $T=\hbar=1$ for simplicity, and hence $J$, $\lambda$, and $V$ are dimensionless quantities. Under periodic boundary conditions~(PBC), Eq.~(\ref{tri}) will be invariant under translation by two lattice sites. As a consequence, we can perform a Fourier transform to express Eq.~(\ref{tri}) in terms of quasimomentum $k$. The reduced Floquet operator can then be written as~(see Appendix~\ref{derive} for details),

\begin{equation}
\mathcal{U}(k,\phi_y,\phi_z)= e^{-\mathrm{i} \left[2J\cos(k)\sigma_x +2\lambda \sin(k)\cos(\phi_y)\sigma_y\right]} e^{-\mathrm{i}V \cos(\phi_z) \sigma_z}\;,
\label{mod}
\end{equation}

\n where it satisfies $U=\sum_k \mathcal{U}(k,\phi_y,\phi_z) \otimes |k\rangle \langle k |$.

In order to realize Eq.~(\ref{tri}) experimentally, we could consider a two-stage process defined as follows in the framework of $1$D lattice. When $2m\leq t < 2m+1$, with $m\in \mathbb{Z}$, we turn off any hopping amplitude between two lattice sites while turning on the on-site potential of the lattice sites given by $V_n=(-1)^n V \cos(\phi_z)$. When $2m+1\leq t<2m+2$, we turn off all the on-site potentials, while turning on the hopping amplitude between two lattice sites given by $t_n=J+(-1)^n \lambda \cos(\phi_y)$. Experimentally, this non-uniform hopping amplitude can be realized by using waveguide arrays in the framework of photonic quasicrystals setup~\cite{exp1}.

\subsection{Analysis of the band touching points}

\begin{figure}
\begin{center}
\includegraphics[scale=0.3]{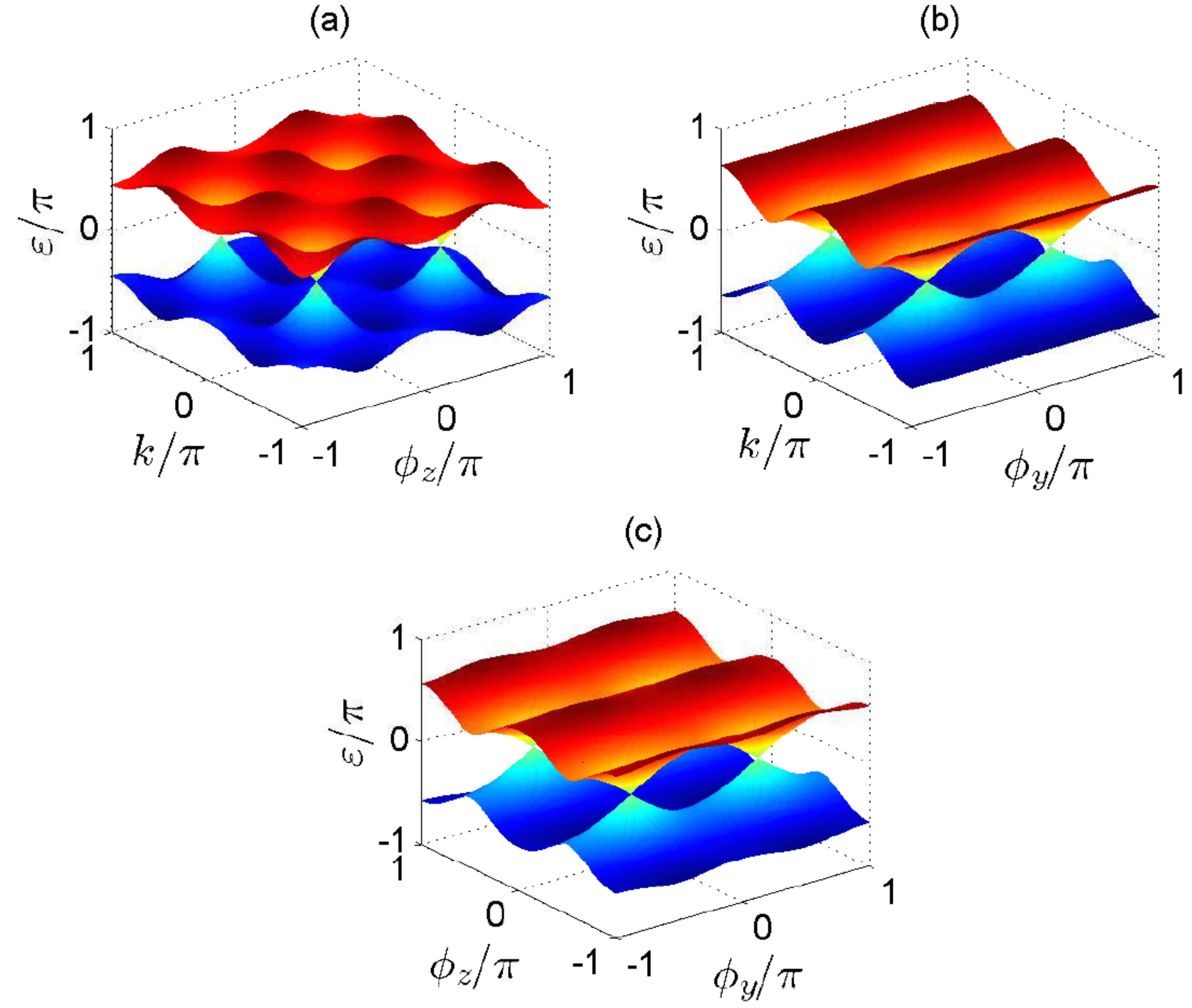}
\end{center}
\caption{(color online). A typical quasienergy spectrum of the ODKHM under PBC at a fixed (a) $\phi_y=\frac{\pi}{2}$, (b) $\phi_z=\frac{\pi}{2}$, and (c) $k=\frac{\pi}{2}$, under $J=1$, $\lambda=0.5$, $V=2$. Here and in all other figures, plotted quantities are in dimensionless units.}
\label{spect1}
\end{figure}

By diagonalizing Eq.~(\ref{mod}) at given values of $k$, $\phi_y$, and $\phi_z$ and scan these parameters over the whole ``Brillouin zone", i.e., $(k,\phi_y,\phi_z)\in [-\pi,\pi)\times [-\pi,\pi)\times [-\pi,\pi)$, we will be able to get an effective quasienergy spectrum of Eq.~(\ref{mod}), where quasienergy $\varepsilon$ is defined as the phase of the eigenvalue of Eq.~(\ref{mod}), satisfying $\mathcal{U}(k,\phi_y,\phi_z)|\psi\rangle=e^{-\mathrm{i}\frac{\varepsilon(k,\phi_y,\phi_z) T}{\hbar}}|\psi\rangle$ with $|\psi\rangle$ be an eigenvector of $\mathcal{U}$. Fig.~(\ref{spect1}) shows the quasienergy spectrum at $J=1$, $\lambda=0.5$, and $V=2$. It suggests that the two bands touch at $(k,\phi_y,\phi_z)=\left(\pm \frac{\pi}{2},\pm \frac{\pi}{2}, \pm \frac{\pi}{2}\right)$. Furthermore, near each of these touching points, the quasienergy dispersion is linear with respect to any two parameters when the other parameter is fixed. In fact, a detailed calculation in the Appendix \ref{exp} shows that by expanding Eq.~(\ref{mod}) around each of these touching points, the effective Hamiltonian is found to be in the form of a Weyl Hamiltonian with chirality $\chi=\pm\mathrm{sgn}(J\lambda V)$~\cite{Hosur}.

\begin{figure}
\begin{center}
\includegraphics[scale=0.3]{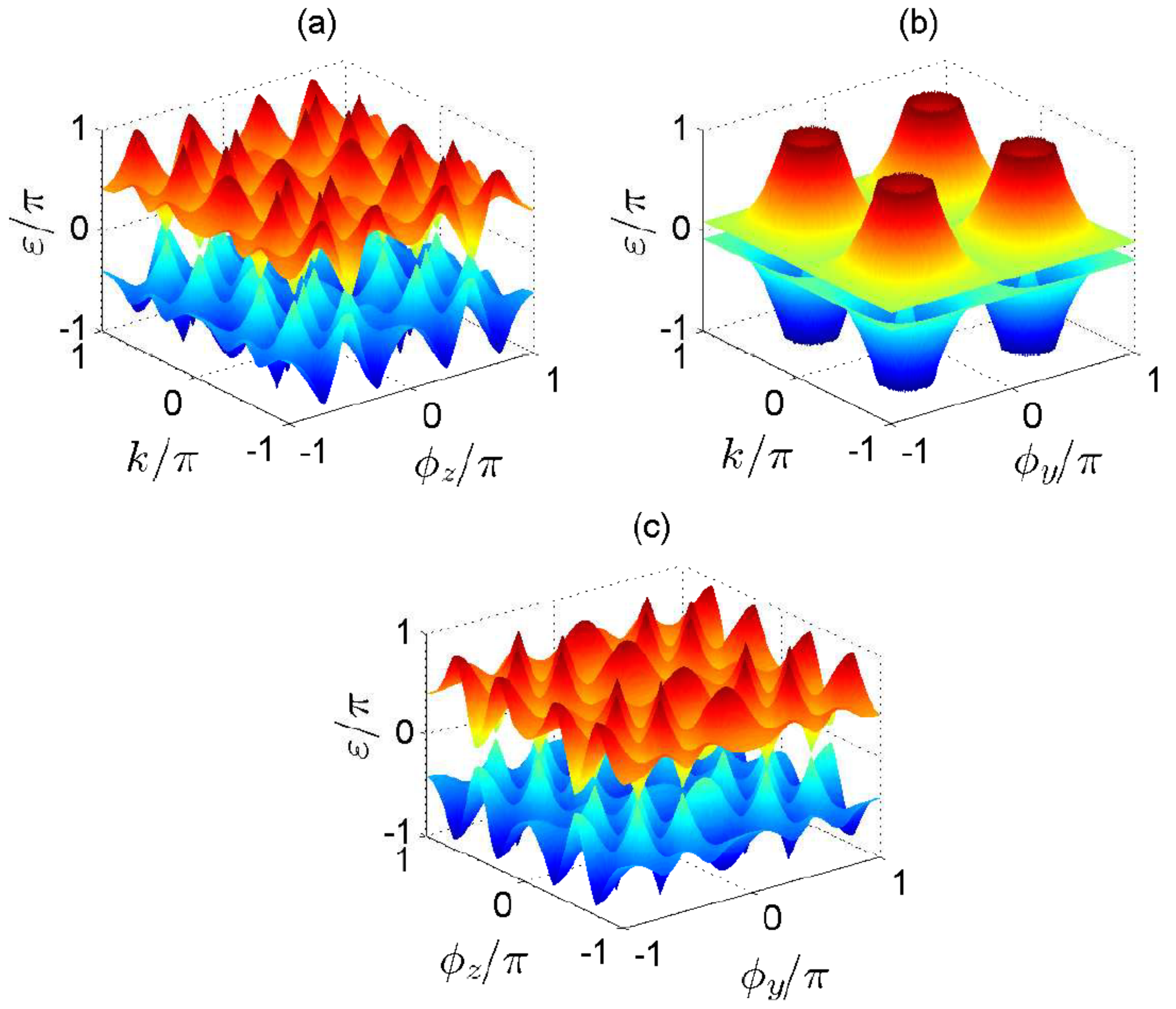}
\end{center}
\caption{(color online). A typical quasienergy spectrum of the ODKHM under PBC at a fixed $\phi_y=\frac{\pi}{2}$ (a), $\phi_z=\frac{\pi}{2}$ (b), and $k=\frac{\pi}{2}$ (c), under $J=3$, $\lambda=3$, $V=5$.}
\label{spect2}
\end{figure}

By increasing the value of parameters $J$, $\lambda$, and $V$, we found that more band touching points start to appear at both quasienergy $0$ and quasienergy $\pm \pi$, as illustrated in Fig.~(\ref{spect2}), an observation which is absent in static systems such as the non-kicked version of the model~\cite{Sarma1, Sarma2}. In order to characterize all the band touching points, we shall first introduce the following Lemma, whose proof is provided in Appendix \ref{proof}.

\begin{lemma}
	Let $\vec{n}$ and $\vec{m}$ be two perpendicular unit vectors, and $a$ and $b$ be functions of three periodic variables $(k,\phi_y,\phi_z)$. The band touching points of the quasienergy spectrum of $e^{\mathrm{i} a \vec{n}\cdot \sigma} e^{\mathrm{i} b \vec{m}\cdot \sigma}$ can only occur if $a$ and $b$ are integer multiples of $\pi$. \label{lem}
\end{lemma}

\n In order to use Lemma \ref{lem}, we need to first turn Eq.~(\ref{mod}) into the following form,

\begin{equation}
\mathcal{U}(k,\phi_y,\phi_z)= e^{-\mathrm{i} f \hat{r}\cdot \sigma} e^{-\mathrm{i}V \cos(\phi_z) \sigma_z}\;,
\label{mod2}
\end{equation}

\n where $f=\sqrt{4J^2\cos^2(k)+4\lambda^2 \cos^2(\phi_y)\sin^2(k)}$, $\hat{r}=\frac{2J}{f} \cos(k)\hat{x}+\frac{2\lambda}{f} \sin(k)\cos(\phi_y)\hat{y}$, and $\hat{x}$, $\hat{y}$, $\hat{z}$ are unit vectors along $x$, $y$, and $z$ directions, respectively. Since $\hat{r} \perp \hat{z}$, Lemma \ref{lem} is applicable, and it suggests that the band touching points appear when $f=l_1 \pi$ and $V\cos(\phi_z)=l_2\pi$, where $l_1, l_2 \in \mathbb{Z}$. All possible band touching points that appear in this model can be classified into three types.

\begin{center}
\subsubsection*{\textbf{Type 1}: $l_1=0$ and $l_2\in \mathbb{N}$.}
\end{center}

Since the quasienergy is only defined modulo $2\pi$, Eq.~(\ref{mod2}) implies that if $l_2$ is even, the two bands touch at quasienergy $0$, while if $l_2$ is odd, they touch at quasienergy $\pi$. Furthermore, $l_1=0$ implies that $f=0$, and therefore $k=\pm\frac{\pi}{2}$ and $\phi_y=\pm\frac{\pi}{2}$. In order for the solutions to exist, $|V|\geq l_2\pi$, so that $\phi_z=\arccos\left(\frac{l_2\pi}{V}\right)$. It is therefore clear from this analysis that new band touching points will start to appear as $V$ is increased by $\pi$. Therefore, by controlling the value of $V$, we will be able to generate as many band touching points of this type as possible.

By expanding Eq.~(\ref{mod}) around one of these band touching points up to first order in $k$, $\phi_y$, and $\phi_z$, we obtain the following,

\begin{equation}
\mathcal{U}(k_x,k_y,k_z)=e^{-\mathrm{i}\left\lbrace l_2\pi- [2J k_x \sigma_x +2\lambda k_y\sigma_y+V\sin(\phi_1)k_z\sigma_z]\right\rbrace}\;,
\label{ham1}
\end{equation}

\n where we have defined $k_x=k-\frac{\pi}{2}$, $k_y=\phi_y-\frac{\pi}{2}$, $k_z=\phi_z-\phi_1$, with $\phi_1=\arccos\left(\frac{l_2\pi}{V}\right)$. The details of this calculation is presented in Appendix \ref{exp}. According to Eq.~(\ref{ham1}), the effective Hamiltonian $h_\mathrm{eff}$ associated with the Floquet operator Eq.~(\ref{mod}), which is defined according to $\mathcal{U}=e^{-\mathrm{i}h_\mathrm{eff}}$, is in the form of a Weyl Hamiltonian with chirality $\chi=-\mathrm{sgn}\left[J\lambda V\sin(\phi_1)\right]$. The analysis therefore shows that this type of band touching points corresponds to a Weyl point. In Sec.~\ref{III} A, we will show that under OBC along the physical dimension, a line connecting two Weyl points associated with this type of band touching points~(Fermi arc) will appear.

\begin{center}
\item \subsubsection*{\textbf{Type 2}: $l_1\neq 0$, $l_2\in \mathbb{N}$, with $J\geq \frac{l_1 \pi}{2}\geq\lambda$.}
\end{center}

By referring to the proof of Lemma \ref{lem} presented in Appendix \ref{proof}, it follows that the two bands touch at quasienergy $\pi$ when either $l_1$ or $l_2$ is odd and the other is even, and they touch at quasienergy $0$ when both $l_1$ and $l_2$ are even or odd. Furthermore, similar to type $1$, in order for the solutions to exist, $|V|\geq l_2\pi$, so that $\phi_z=\arccos\left(\frac{l_2\pi}{V}\right)$. However, since $l_1\neq 0$, $k$ and $\phi_y$ can in general take a continuous range of values satisfying $f=l_1\pi$. By treating $\phi_y$ as an independent variable, the two bands touch if $k$ satisfies the following condition,

\begin{equation}
k=\pm \arccos\left[\sqrt{\frac{l_1^2 \pi^2- 4\lambda^2 \cos(\phi_y)^2}{4J^2-4\lambda^2\cos(\phi_y)^2}}\right]\;.
\label{line}
\end{equation}

\begin{figure}
\begin{center}
\includegraphics[width=0.5\textwidth]{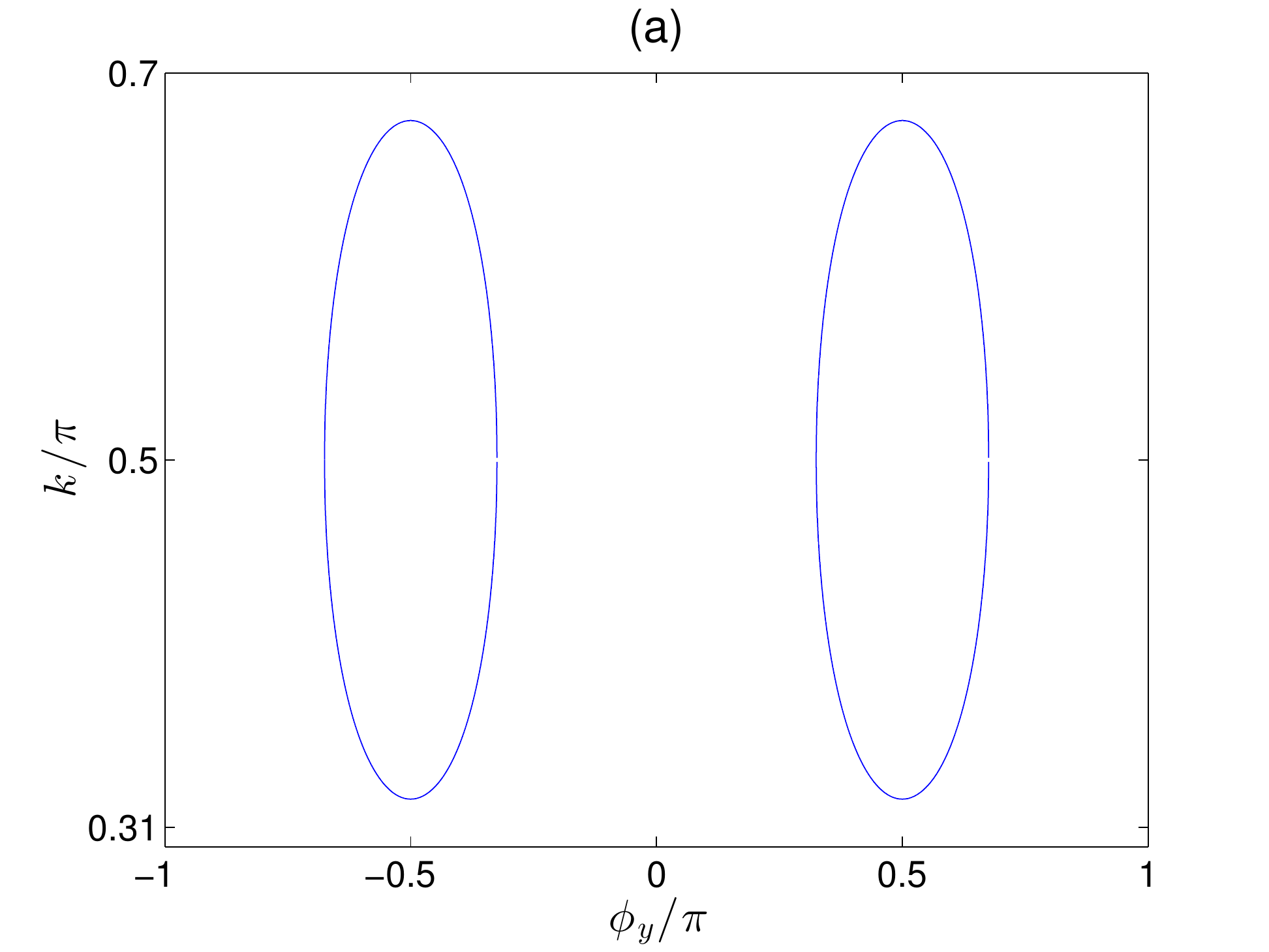}%
\includegraphics[width=0.5\textwidth]{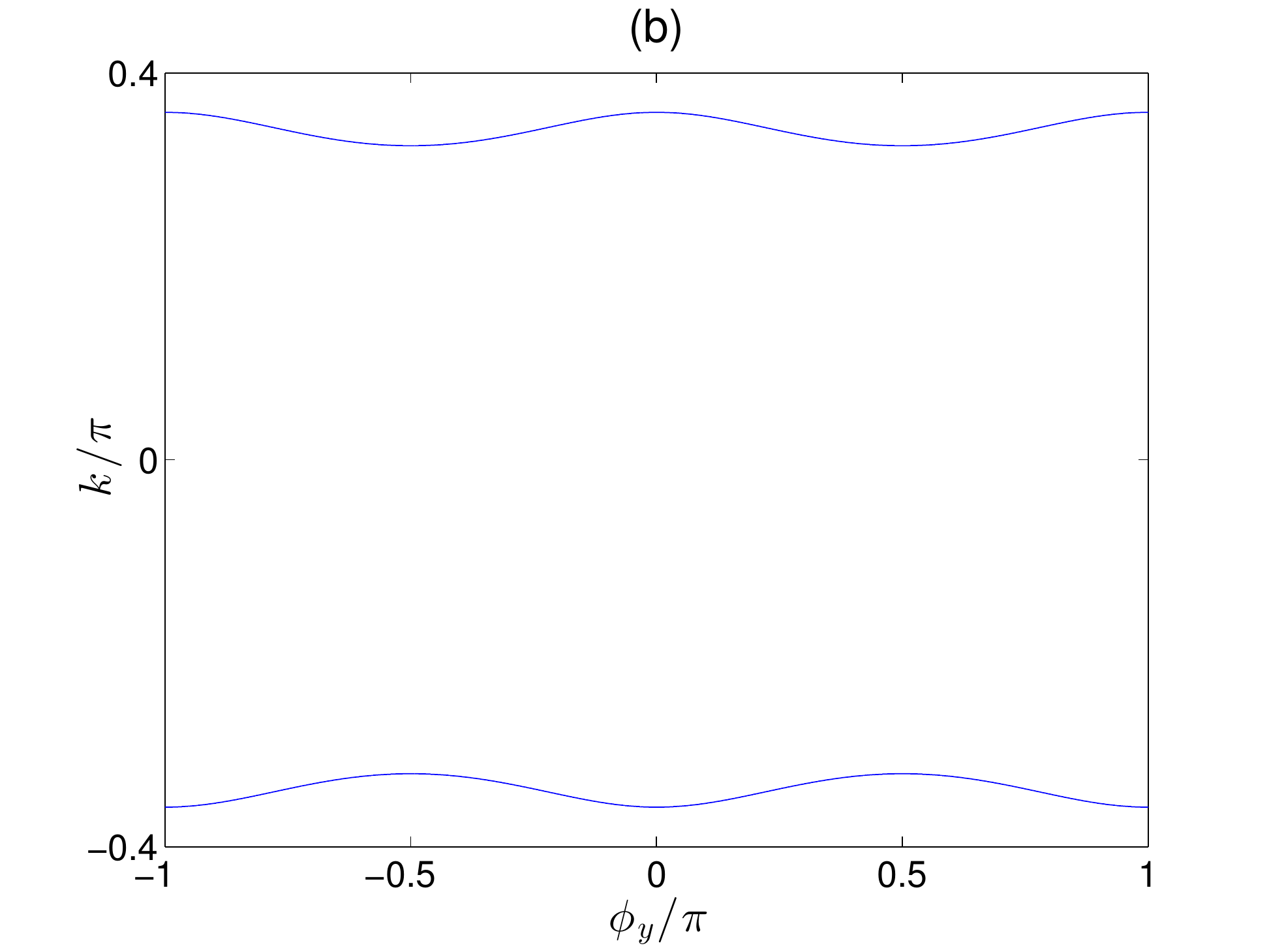}
\end{center}
\caption{ A typical shape of (a) type 3 and (b) type 2 line nodes. The plot is taken at $J=3$, $V=2$, (a) $\lambda=3$ and (b) $\lambda=1$.}
\label{type23}
\end{figure}

Since $J\geq \frac{l_1\pi}{2}\geq \lambda $, a solution to $k$ exists for any $\phi_y\in [-\pi,\pi)$. As a result, the two bands will touch along a curve as a function of $\phi_y$ and $k$. Fig.~\ref{type23}(b) shows a typical example of the line nodes of this type at $J=3$, $\lambda=1$, and $V=2$, i.e., $l_1=1$ and $l_2=0$. By expanding Eq.~(\ref{mod}) around $(k,\phi_y,\phi_z)=\left(k_0',\frac{\pi}{2},\frac{\pi}{2}\right)$, where $k_0'=\arccos\left(\frac{\pi}{2J}\right)$, which corresponds to a point where the two bands touch, up to first order in $k$, $\phi_y$, and $\phi_z$, the effective Hamiltonian is found to be~(See Appendix \ref{exp} for the calculation details),

\begin{equation}
\begin{split}
&h_\mathrm{eff}(k_x,k_y,k_z)=\left[\pi -\Phi(k_x,k_z)\right]\hat{n}\cdot \sigma +\frac{2J k_x \sin(k_0')}{\Phi(k_x,k_z)}\left[-2 \lambda k_y\sin(k_0')+\pi V k_z \right] \sigma_y \;,
\end{split}
\label{lineff}
\end{equation}

\n where $k_x=k-k_0'$, $k_y=\phi_y-\frac{\pi}{2}$, $k_z=\phi_z-\frac{\pi}{2}$, $\Phi(k_x,k_z)=\sqrt{V^2 k_z^2+4J^2 k_x^2 \sin^2(k_0')}$, and $\hat{n}=-\frac{Vk_z}{\Phi(k_x,k_z)}\hat{z}+\frac{2J k_x \sin(k_0')}{\Phi(k_x,k_z)}\hat{x}$. The quasienergy associated with this effective Hamiltonian is $\varepsilon \approx \pm \left[\pi - \Phi(k_x,k_z)\right]$.

According to Ref.~\cite{Burkov}, line nodes can be topologically characterized by a topological winding number, which is constructed as follows. On $k_z = \frac{2\lambda \sin(k_0')}{\pi V}k_y$ plane, the effective Hamiltonian, as defined in type $1$, depends only on two Pauli matrices and can be written in the following form,

\begin{equation}
h_\mathrm{eff}(k_x,k_z)=h_1(k_x, k_z) \sigma_x+h_2(k_x,k_z) \sigma_z \;.
\end{equation}

\n Define a complexified Hamiltonian $h=h_2+\mathrm{i}h_1=|h|e^{\mathrm{i} \phi}$. The topological winding number associated with this line node is defined as,

\begin{equation}
n=\frac{1}{2\pi}\oint d\mathbf{k}\cdot \nabla_\mathbf{k} \phi \;,
\end{equation}

\n which can in general take values $0$ or $\pm 1$~(where the sign depends on the direction of the line integral), and is closely related to the Berry phase around a loop surrounding the line node~\cite{Burkov}. By considering a loop parametrized by $\theta$ according to $k_x = r \sin(\theta)$ and $k_z=\frac{r\pi V}{\sqrt{\pi^2 V^2+4\lambda^2 \sin^2(k_0')}} \cos(\theta)$, it follows that $\tan(\phi) = -\frac{2J\sin(k_0')\sqrt{\pi^2 V^2+4\lambda^2 \sin^2(k_0')}}{\pi V^2} \tan(\theta)$. This suggests that as $\theta$ goes from $0$ to $2\pi$, $\phi$ goes from $0$ to $-2\pi$, so that $\phi(2\pi)=-2\pi$ and $\phi(0)=0$. Therefore,

\begin{equation}
n = \frac{1}{2\pi} \int_0^{2\pi} d\theta \frac{d \phi}{d\theta} = \frac{1}{2\pi} \left[\phi(2\pi)-\phi(0)\right]= - 1\;,
\end{equation}

\n which shows that the line nodes of this type are topologically nontrivial.

\begin{center}
\item \subsubsection*{\textbf{Type 3:} $l_1\neq 0$, $l_2\in \mathbb{N}$, with $\lambda \geq \frac{l_1 \pi}{2}$.}
\end{center}

Similar to type $2$, the two bands touch at quasienergy $0$ when both $l_1$ and $l_2$ are even or odd, and at quasienergy $\pi$ when either $l_1$ or $l_2$ is even, while the other is odd. Also, similar to both type $1$ and $2$, in order for the solutions to exist, $|V|\geq l_2\pi$, so that $\phi_z=\arccos\left(\frac{l_2\pi}{V}\right)$. By treating $\phi_y$ as an independent variable, $k$ will need to satisfy Eq.~(\ref{line}) in order for the two bands to touch. However, since a cosine function can only take values between $-1$ and $1$, a solution to $k$ will only exist at some values of $\phi_y$ in $[-\pi,\pi)$. If $J\geq \frac{l_1 \pi}{2}$, a solution to $k$ exists as long as the numerator of Eq.~(\ref{line}) is $l_1^2 \pi^2- 4\lambda^2 \cos^2(\phi_y)\geq 0$, or $\phi_y \in [-\pi+\phi_0,-\phi_0]\cup [\phi_0, \pi-\phi_0]$, where $\phi_0=\arccos\left(\frac{l_1\pi}{2\lambda}\right)$. If $J\leq \frac{l_1 \pi}{2}$, a solution to $k$ exists as long as the numerator of Eq.~(\ref{line}) is $l_1^2\pi^2-4\lambda^2\cos^2(\phi_y)\leq 0$. In both cases, the two bands will touch along a closed curve within $k$ and $\phi_y$ Brillouin zone. As an illustration, Fig.~\ref{type23}(a) shows a typical line node of this type at $J=3$, $\lambda=3$, and $V=2$, i.e., $l_1=1$ and $l_2=0$.

By expanding Eq.~(\ref{mod}) around $(k,\phi_y,\phi_z)=\left(\frac{\pi}{2},\phi_0,\frac{\pi}{2}\right)$ up to first order in $k$, $\phi_y$, and $\phi_z$, i.e., a point where the two bands touch, the effective Hamiltonian is found to be,

\begin{equation}
h_\mathrm{eff} = \left[\pi -\Phi'(k_y,k_z)\right]\hat{m}\cdot \sigma -\frac{2\lambda k_y \sin(\phi_0)}{\Phi'(k_y,k_z)}\left( \pi V k_z+ 2 J k_x\right) \sigma_x \;,
\label{lineff2}
\end{equation}

\n where $k_x=k-\frac{\pi}{2}$, $k_y=\phi_y-\phi_0$, $k_z=\phi_z-\frac{\pi}{2}$, $\Phi'(k_y,k_z)=\sqrt{V^2 k_z^2+4\lambda^2 k_y^2 \sin^2(\phi_0)}$, and $\hat{m}=\frac{2\lambda k_y \sin(\phi_0)}{\psi(k_y,k_z)}\hat{y}-\frac{Vk_z}{\psi(k_y,k_z)}\hat{z}$. The quasienergy associated with this effective Hamiltonian is $\varepsilon \approx \pm \left[\pi - \Phi'(k_y,k_z)\right]$.

We can also calculate the topological winding number in a similar fashion as in type $2$. On $k_z=-\frac{2J}{\pi V}k_x$ plane, the effective Hamiltonian depends only on two Pauli matrices which can be written as,

\begin{equation}
h_\mathrm{eff}(k_y,k_z)=h_1(k_y, k_z) \sigma_y+h_2(k_y,k_z) \sigma_z \;.
\end{equation}

\n As before, we define a complexified Hamiltonian $h=h_1+\mathrm{i}h_2=|h|e^{\mathrm{i}\phi}$. By parametrizing $k_y=r\cos(\theta)$ and $k_z=-\frac{r\pi V}{\sqrt{\pi^2V^2+4J^2}}\sin(\theta)$, it follows that $\tan(\phi)=\frac{\pi V^2}{2\lambda \sin(\phi_0)\sqrt{\pi^2V^2+4J^2}}\tan(\theta)$. The topological winding number can then be shown to be $n= 1$, which shows that the line nodes of this type are also topologically nontrivial.

\vspace{1cm}

\section{Edge states}\label{III}

One characterized signature of a topological phase is the existence of localized edge states when the system takes OBC. For example, the band structure of a 2D quantum Hall insulator supports localized chiral edge states which traverse the gap between two bands~\cite{QHI, QHI2, QHI3}, while that of a spin Hall insulator supports localized helical edge states~\cite{SHI, SHI2}. Localized edge states can also appear in gapless systems, e.g. Fermi arc in Weyl semimetal~\cite{Burkov, Ferarc, Hosur2, Ferarc2} and surface states in nodal line semimetal~\cite{Burkov, Mullen, LN2, Youngkuk}, which indicates that such systems can also be topologically characterized. We have shown in section \ref{II} that the ODKHM hosts a number of Weyl points and topologically non-trivial line nodes at quasienergy $0$ and quasienergy $\pi$. In this section, we will investigate the emergence of localized edge states associated with these Weyl points and line nodes. The results are hoped to motivate more studies of bulk-edge correspondence in Floquet systems.

\subsection{Edge states under OBC along the physical dimension}

\begin{figure}
\begin{center}
\includegraphics[width=0.5\textwidth]{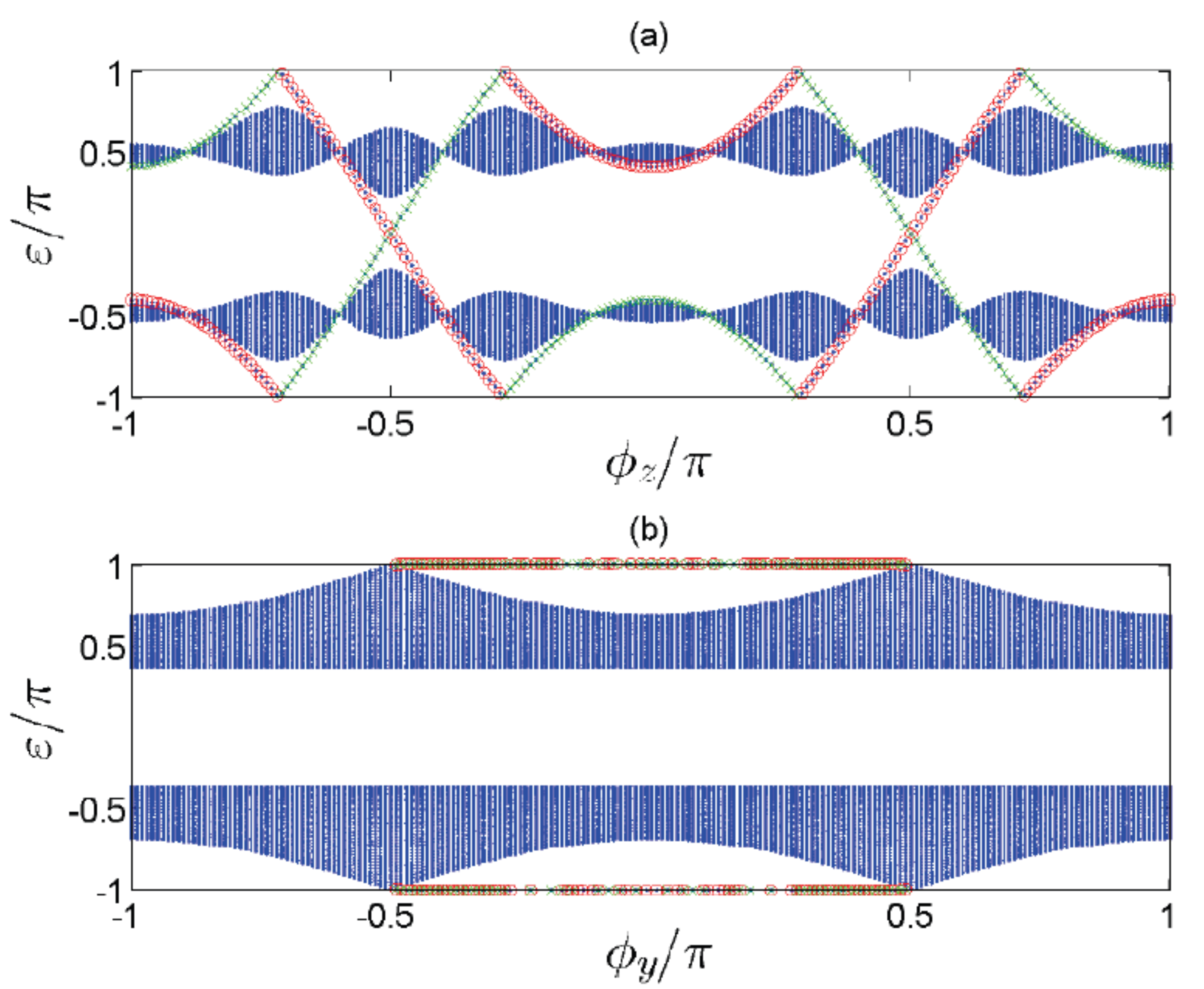}%
\includegraphics[width=0.6\textwidth]{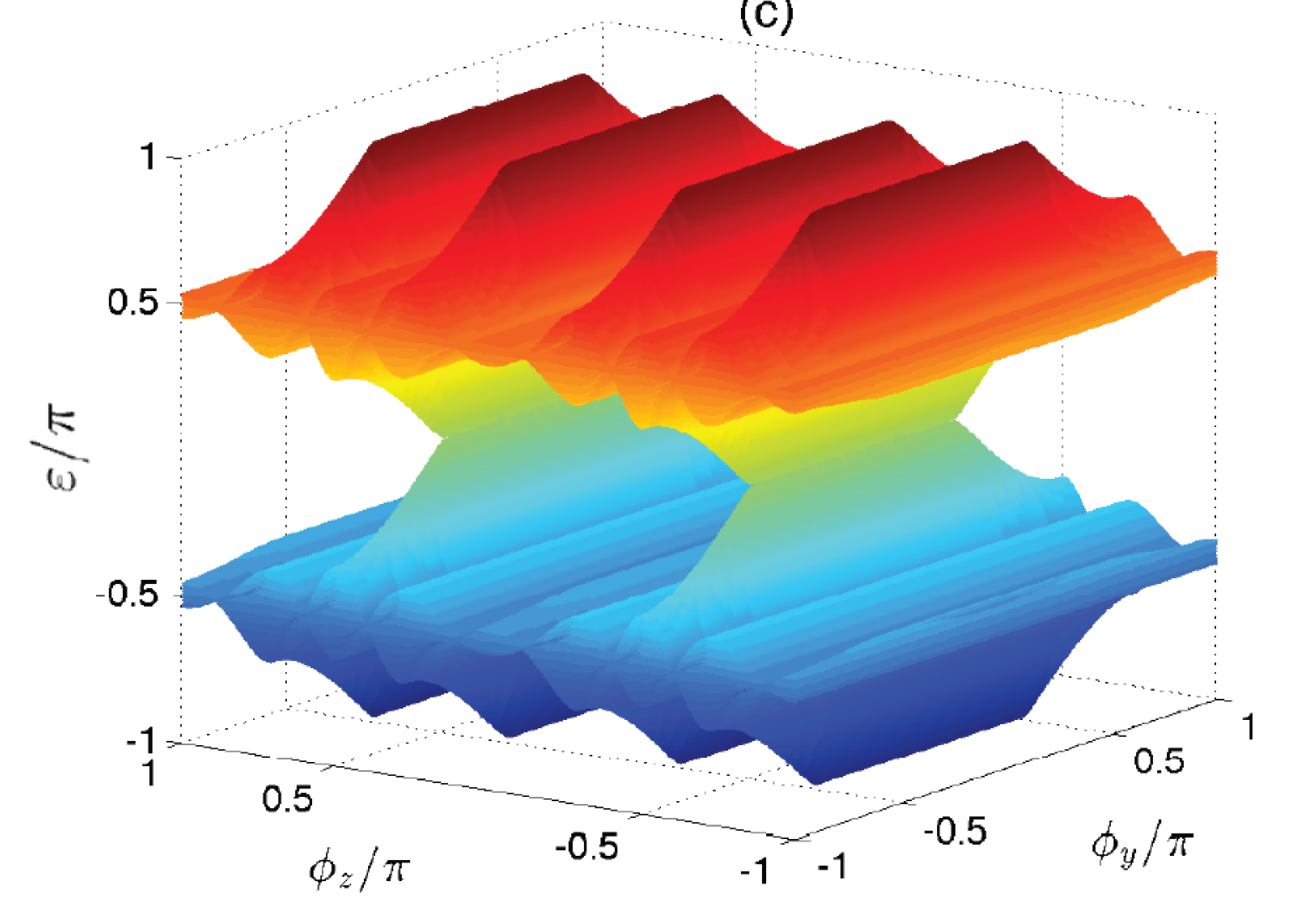}
\end{center}
\caption{(color online). Quasienergy spectrum of Eq.~(\ref{tri}) under OBC at a fixed (a) $\phi_y=\frac{\pi}{4}$ and (b) $\phi_z=\arccos\left(\frac{\pi}{V}\right)$. Red and green marks denote edge states localized at the right and left edges, respectively. The full quasienergy spectrum is presented in (c). The parameters used are $N=100$, $\lambda=0.5$, $J=1$, and $V=5$.}
\label{2}
\end{figure}

Fig.~\ref{2} shows the quasienergy spectrum of the Floquet operator Eq.~(\ref{tri}) under OBC with $N=100$, $\lambda=0.5$, $J=1$, and $V=5$. Under the given parameters, it confirms our analysis in section \ref{II} that there exists some Weyl points at both quasienergy $0$ and quasienergy $\pi$, corresponding to $l_2=0$ and $l_2=1$. A slice of the full quasienergy spectrum at a fixed $\phi_z=\arccos\left(\frac{\pi}{V}\right)$ as shown in Fig.~\ref{2}(b) shows the existence of a line connecting two Weyl points with opposite chiralities, which is analogous to the Fermi arc in static Weyl semimetal systems. The emergence of this Fermi arc shares a similar topological origin as the edge states of the Su-Schrieffer-Heeger~(SSH) model~\cite{SSH}. At a fixed $\phi_z=\arccos\left(\frac{\pi}{V}\right)$, the effective Hamiltonian takes the form,

\begin{equation}
h_\mathrm{eff}=\pi+2J\cos(k)\sigma_x+2\lambda \cos(\phi_y)\sin(k)\sigma_y \;,
\label{inv}
\end{equation}

\n which can be mapped to a static SSH model in the form of $h_\mathrm{SSH}=\pi +\sum_n [(J+\delta t_n)|n\rangle \langle n+1| +h.c.]$, where $\delta t_n=(-1)^n\lambda \cos(\phi_y)$, by a suitable Fourier transform. SSH model can be characterized by a $\mathbb{Z}_2$ topological invariant. It is topologically nontrivial when $\delta t_1$ and $\delta t_{N-1}$ are less than $0$, while it is topologically trivial when $\delta t_1$ and $\delta t_{N-1}$ are greater than $0$. In this case, the signs of $\delta t_1$ and $\delta t_{N-1}$ are controlled by the parameter $\phi_y$. This explains why edge states exist only when $-\frac{\pi}{2}<\phi_y<\frac{\pi}{2}$, and thus appear to form a line connecting two Weyl points at $\phi_y=\pm \frac{\pi}{2}$.

Fig.~\ref{2}(a) shows a slice of the full quasienergy spectrum at a fixed $\phi_y=\frac{\pi}{4}$, where we observe counter-propagating chiral edge states joining the two bands. These counter-propagating edge states exist for any fixed value of $\phi_y$ in $(-\frac{\pi}{2},\frac{\pi}{2})$. When $|\phi_y|=\frac{\pi}{2}$, the two gaps at quasienergy $0$ and $\pi$ close. By increasing $|\phi_y|$ further, the gaps reopen and the edge states disappear. Furthermore, for any value of $\phi_y$ where the edge states appear, they cross at $\phi_z=\left\lbrace \pm\frac{\pi}{2}, \pm \arccos\left(\frac{\pi}{V}\right), \pm\left[\pi-\arccos\left(\frac{\pi}{V}\right)\right]\right\rbrace$, where the effective Hamiltonian has inversion symmetry and can be written in the form of Eq.~(\ref{inv}). These edge states crossings are what appears as a Fermi arc joining two Weyl points.

\begin{figure}
\begin{center}
\includegraphics[width=0.5\textwidth]{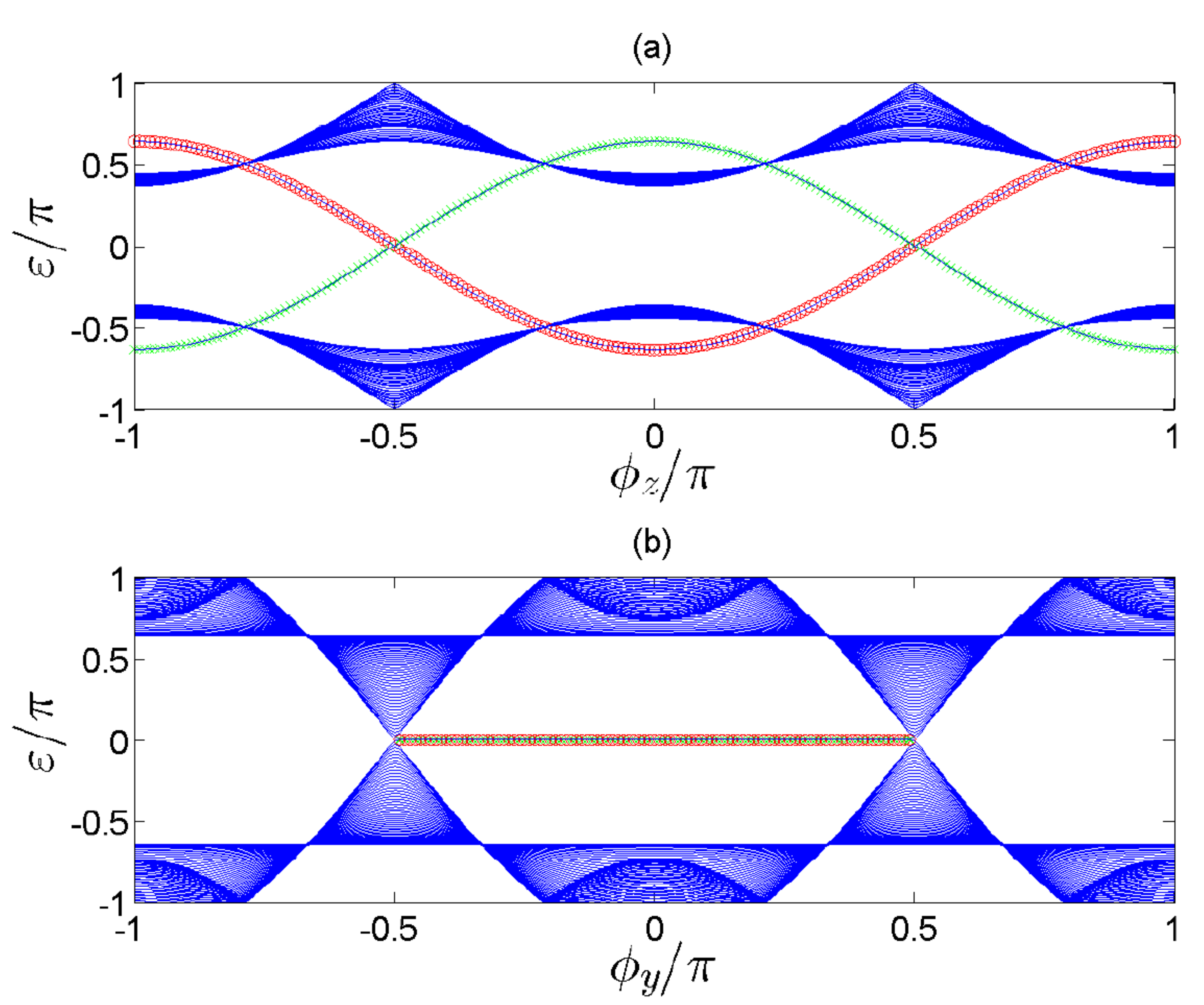}%
\includegraphics[width=0.5\textwidth]{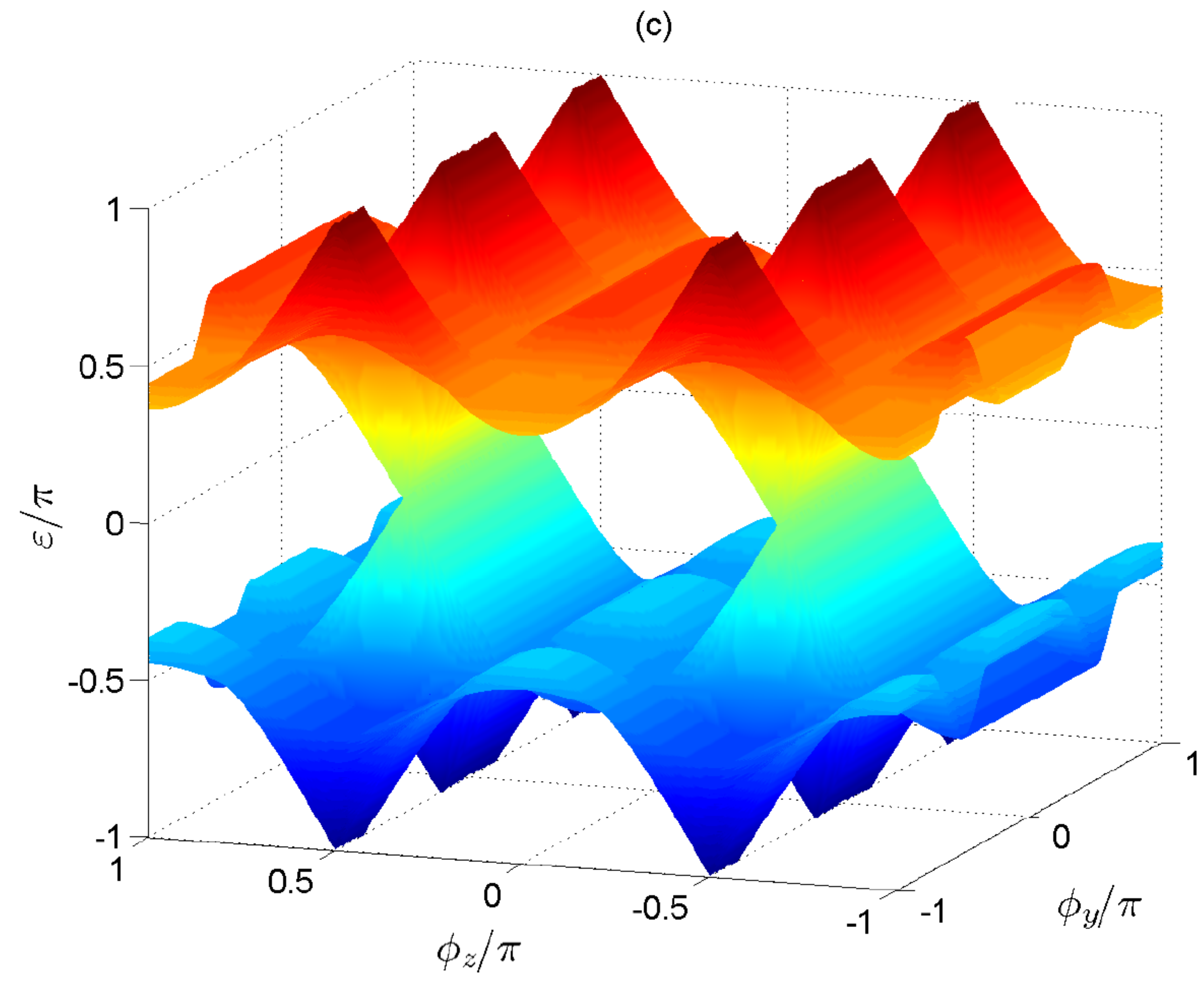}
\end{center}
\caption{(color online). Quasienergy spectrum of Eq.~(\ref{tri}) under OBC at a fixed (a) $\phi_y=0$ and (b) $\phi_z=\frac{\pi}{2}$. Red and green marks denote edge states localized at the right and left edges, respectively. The full quasienergy spectrum is presented in (c). The parameters used are $N=100$, $\lambda=2$, $J=1$, and $V=2$.}
\label{3}
\end{figure}

Fig.~\ref{3} shows another quasienergy spectrum of Eq.~(\ref{tri}) under OBC with $J=1$, $\lambda=2$, and $V=2$, where the line nodes are predicted to appear. Apart from the counter-propagating edge states and the Fermi arcs we mentioned earlier, there does not seem to exist any additional edge states arising as a consequence of the line nodes. Therefore, in order to be able to capture the topology of the line nodes, we take a look at the system under OBC along an artificial dimension, which will be presented in section \ref{III} B.

\subsection{Edge states under OBC along an artificial dimension}

\begin{figure}
\begin{center}
\includegraphics[width=0.5\textwidth]{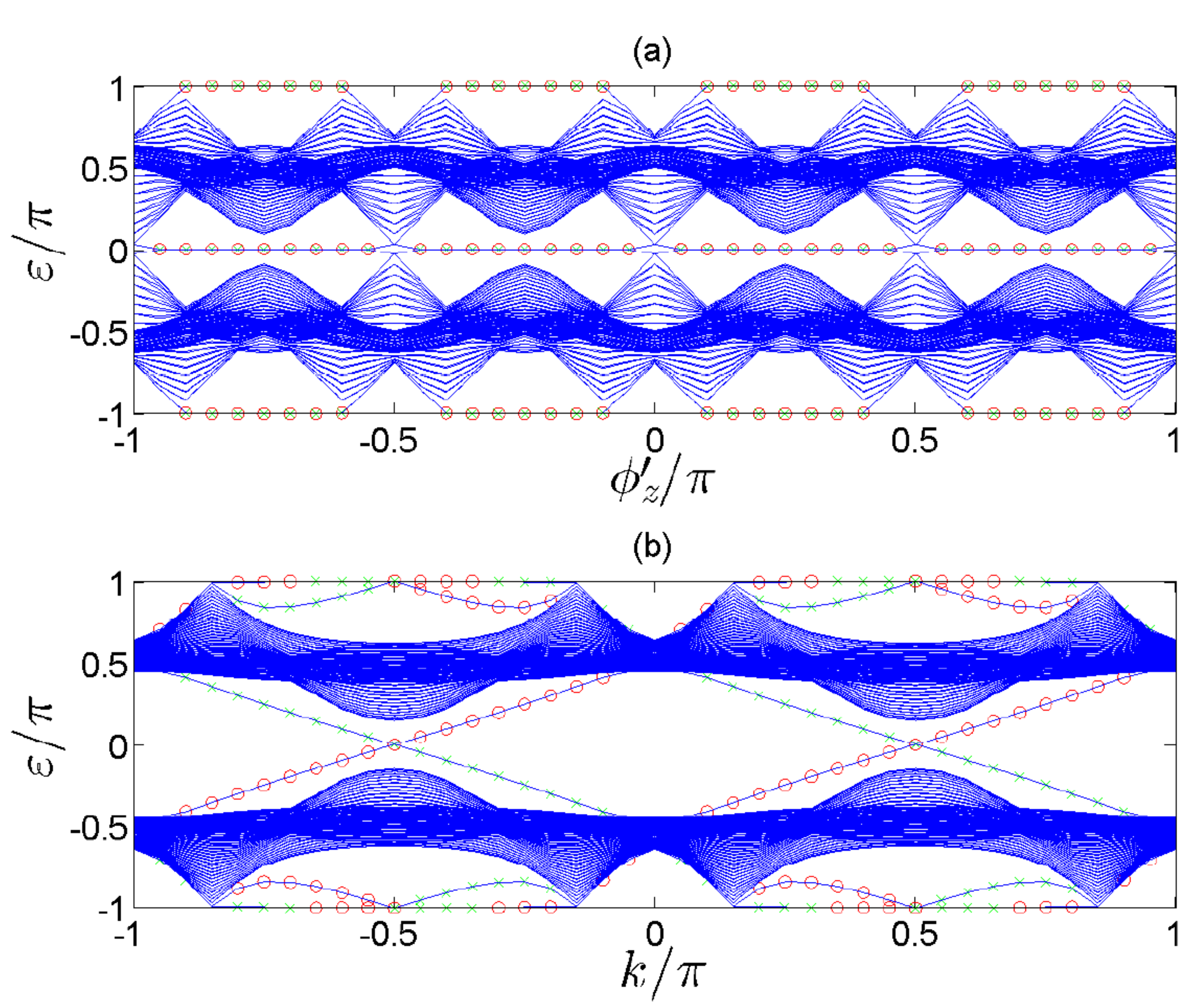}%
\includegraphics[width=0.5\textwidth]{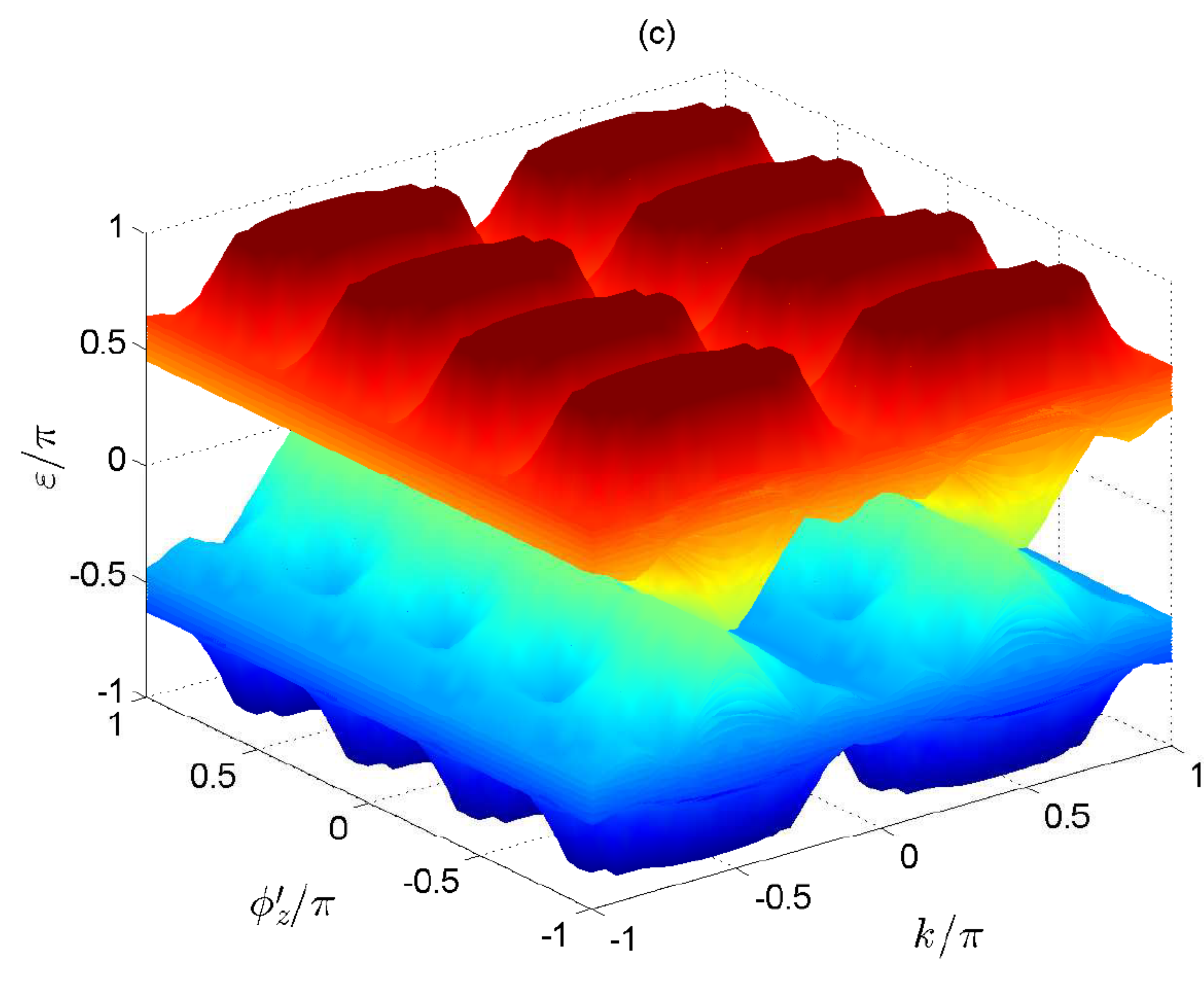}
\end{center}
\caption{(color online). Quasienergy spectrum of Eq.~(\ref{uy}) under OBC at a fixed (a) $k=\frac{\pi}{2}$ and (b) $\phi_z'=\frac{\pi}{5}$. Red and green marks denote edge states localized at the right and left edge, respectively. The full quasienergy spectrum is presented in (c). The parameters used are $N_y'=100$, $\lambda=3$, $J=1$, and $V=2$.}
\label{4}
\end{figure}

OBC along an artificial dimension may be experimentally realized by decoupling the atomic internal degrees of freedom of the lattice that generate the phase shift $\phi_y$ or $\phi_z$~\cite{Art1, Art2}. Mathematically, this can be done by performing a Fourier transform to Eq.~(\ref{mod}) with respect to $\phi_y$ or $\phi_z$ to get an effective lattice model, then taking OBC along this dimension by considering a finite number of lattice sites. Here, the Fourier transformed lattice site corresponding to $\phi_y$ and $\phi_z$ can be imagined as $y$ and $z$ directions, respectively. We found that even under OBC along $y$ or $z$ direction, there will still be no additional edge states associated with the line nodes. However, by opening the system along a diagonal direction in $y$ and $z$, edge states associated with the line nodes become visible, as shown in Fig.~\ref{4}(c).

To be more explicit, we define $\phi_y'=\frac{\phi_y+\phi_z}{2}$ and $\phi_z'=\frac{\phi_z-\phi_y}{2}$. By performing a Fourier transform with respect to $\phi_y'$, Eq.~(\ref{mod}) can be transformed into the following form,

\begin{equation}
U_{y'} = e^{-\mathrm{i} \sum_{n_y'=1}^{N_y'-1} \left[J\cos(k) \sigma_x \otimes |n_y' \rangle \langle n_y' | +\frac{\lambda}{2}\sin(k) e^{\mathrm{i}\phi_z'}\sigma_y\otimes |n_y'+1 \rangle \langle n_y' |+h.c. \right] } e^{-\mathrm{i} \sum_{n_y'=1}^{N_y'-1}\left[\frac{V}{2}e^{-\mathrm{i}\phi_z'}\sigma_z\otimes |n_y'+1 \rangle \langle n_y' | +h.c.\right]}\;,
\label{uy}
\end{equation}

\n where $n_y'$ represents a lattice site along the diagonal direction. Fig.~\ref{4}(c) is then obtained by plotting the quasienergy spectrum of Eq.~(\ref{uy}) with $N_y'=100$, $J=1$, $\lambda=3$, and $V=2$, i.e., when line nodes of type $3$ are expected to appear. At a fixed $k=\frac{\pi}{2}$, Fig.~\ref{4}(a) shows the emergence of degenerate $\pi$-modes inside each line node. These degenerate $\pi$-modes are very similar to the Fermi arc, since they correspond to the crossing between two edge states localized at opposite edges, which is illustrated in Fig.~\ref{4}(b).

\begin{figure}
\begin{center}
\includegraphics[scale=0.2]{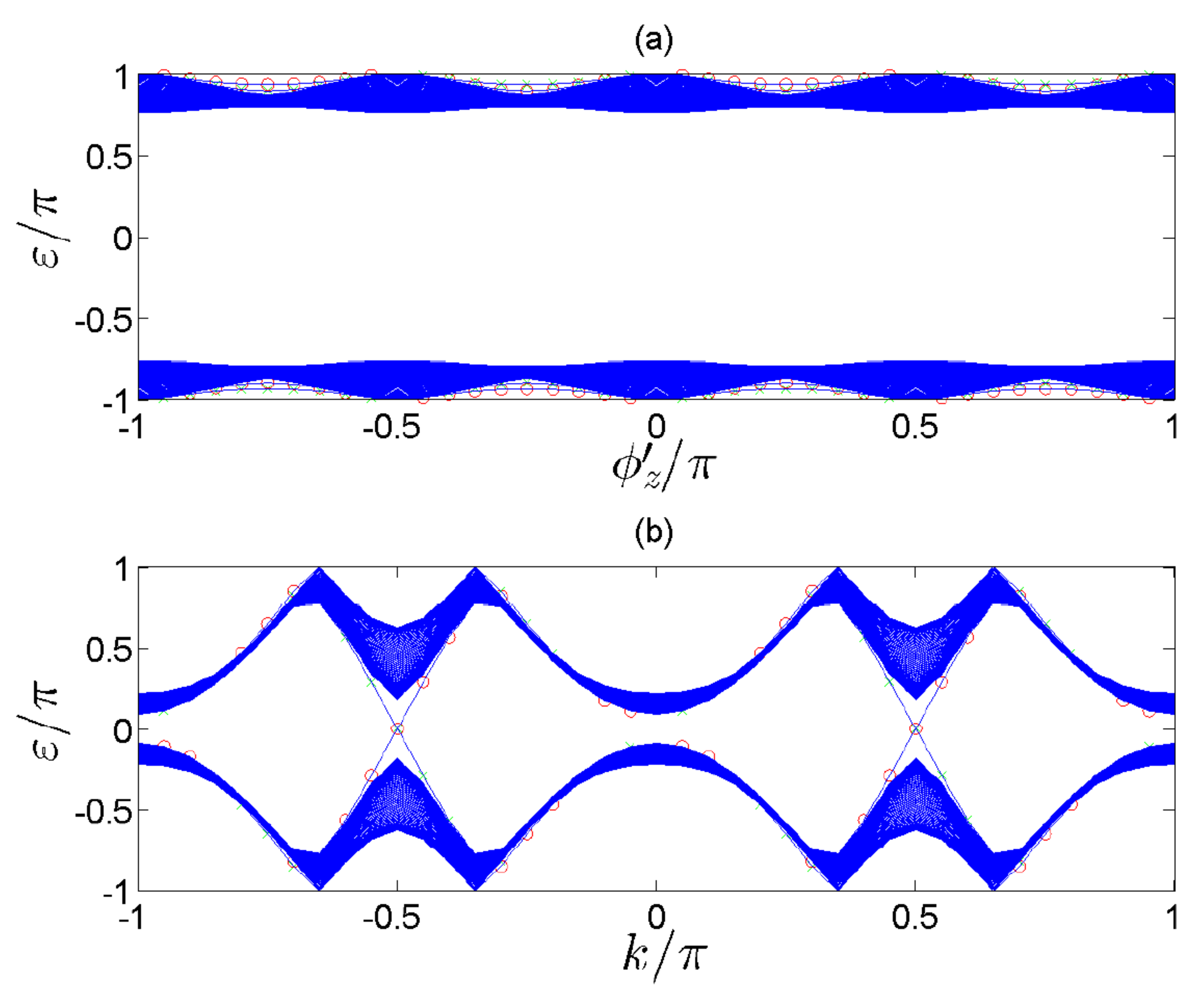}
\end{center}
\caption{(color online). Quasienergy spectrum of Eq.~(\ref{uy}) under OBC at a fixed (a) $k=\arccos\left(\frac{\pi}{6}\right)$ and (b) $\phi_z'=1$. Red and green marks denote edge states localized at the right and left edge, respectively. The parameters used are $N_y'=100$, $J=3$,  $\lambda=\frac{1}{\arccos\left(\frac{\pi}{6}\right)}$, and $V=\frac{2}{\pi}$.}
\label{5}
\end{figure}

As another illustration, Fig.~\ref{5}(a) and \ref{5}(b) show the quasienergy spectrum of Eq.~(\ref{uy}) under OBC at another set of values of the parameters $J$, $\lambda$, and $V$, where line nodes of type $2$ are expected to appear, at a fixed $k=\arccos\left(\frac{\pi}{6}\right)$ and $\phi_z'=1$, respectively. The edge states appearing from a point on a line node and disappearing at another point on the same line node are visible, albeit not as clear as the emergence of edge states when line nodes of type $3$ appear. Nevertheless, the existence of these edge states confirms the fact that both line nodes of type $2$ and type $3$ have topological origins, although this might be associated with a weak topological effect~\cite{WTI, WTI2, WTI3}, as the emergence of edge states depends on the direction in which we ``cut" the system.

\section{Dynamics studies}\label{IV}

Another way to capture the Floquet band topology is through the quantization of particle transport by means of an adiabatic pumping protocol~\cite{Thouless, Derek2}. In this section, we will show that the topology of the Weyl points and line nodes appearing in the ODKHM can be manifested by designing certain adiabatic pumping schemes, which are relatively easy to realize experimentally since they only involve tuning the experimental parameters $\phi_y$ and $\phi_z$ without the need to accelerate the particle by changing $k$.

\subsection{Quantized adiabatic pumping around Weyl points}

According to Ref.~\cite{Hosur}, the chirality characterizing a Weyl point represents a source of Berry curvature in quasimomentum space. It means that if we form a closed $2$D surface enclosing a Weyl point in quasimomentum space parametrized by two parameters $k_1$ and $k_2$, the Chern number associated with the Brillouin zone formed by $k_1$ and $k_2$ equals the chirality of the Weyl point enclosed.

Therefore, in order to capture the chirality of a Weyl point, we first prepare an initial state as a uniform superposition of the lower-band Floquet eigenstates over all $k$ in the Brillouin zone~(Wannier state centered around $n=0$)~\cite{foot1}. That is,

\begin{equation}
| \Psi_0 \rangle = \frac{a}{2\pi}\int_{-\frac{\pi}{a}}^\frac{\pi}{a} |\psi_{-,k}(\phi_y,\phi_z) \rangle dk \;,
\label{is}
\end{equation}

\n where $a$ is the lattice constant. Next, an adiabatic pumping protocol is realized by tuning $\phi_y$ and $\phi_z$ adiabatically according to $\phi_y=\phi_{y,0}+r\sin[\theta(t)]$ and $\phi_z=\phi_{z,0}+r\cos[\theta(t)]$, where $\theta(t)$ is a periodic function of time with a period much longer than the kicking period. In this work, we consider a discrete linear sweeping protocol by changing $\theta$ by $\frac{2\pi}{N}$ every time step $\Delta t=1$, i.e., $\theta(t)=\frac{2\pi i}{N}$ for $i-1<t\leq i$ and $0<i\leq N$. Our objective is to evaluate the shift in position expectation value after one adiabatic cycle $t=N$.

Analytically, Eq.~(\ref{chiral}) derived in Appendix \ref{quantized} implies the following two observations:

\begin{enumerate}
\item The change in position expectation value of the wave packet~($\Delta \langle X\rangle$) after one adiabatic cycle is quantized by the total chirality of the Weyl points enclosed.
\item Under the same initial state and adiabatic pumping scheme, the chirality of the Weyl points at quasienergy $\pi$ and quasienergy $0$ contribute to $\Delta \langle X \rangle$ in an opposite manner.
\end{enumerate}

\begin{figure}
\begin{center}
\includegraphics[scale=0.5]{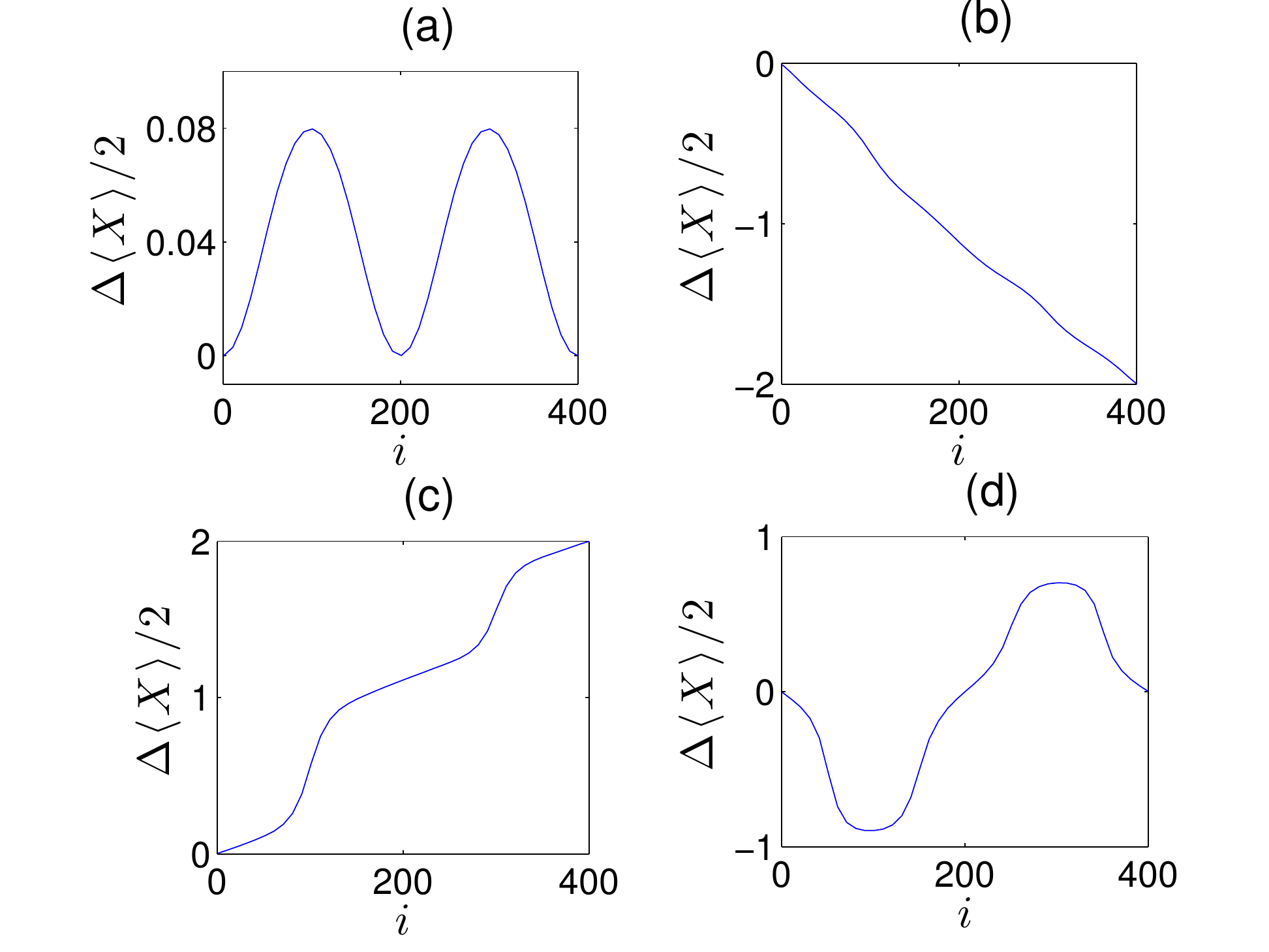}
\end{center}
\caption{The shift in position expectation value after one adiabatic cycle along a loop in the artificial dimensions that (a) does not enclose any Weyl point, (b) encloses two Weyl points at quasienergy $0$, (c) encloses two Weyl points at quasienergy $\pi$ and (d) encloses two Weyl points at quasienergy $0$ and two other Weyl points at quasienergy $\pi$. The parameters used are $J=1$, $\lambda=0.5$, $V=2$ for (a) and (b) and $V=5$ for (c) and (d).}
\label{pump}
\end{figure}

\n The above observations can also be verified numerically, as shown in Fig.~\ref{pump}. Fig.~\ref{pump}(a) is obtained by setting $\phi_{y,0}=\phi_{z,0}=0$ and $r=\frac{\pi}{4}$ so that the adiabatic loop does not enclose any Weyl points, and $\Delta \langle X \rangle$ after one adiabatic cycle is zero. In Fig.~\ref{pump}(b), $\phi_{y,0}=\phi_{z,0}=\frac{\pi}{2}$ and $r=\frac{\pi}{4}$ so that the adiabatic loop encloses two Weyl points at quasienergy $0$ at $(k,\phi_y,\phi_z)=\left(\pm\frac{\pi}{2},\frac{\pi}{2},\frac{\pi}{2}\right)$, and $\frac{\Delta\langle X \rangle}{2a}=-2$~\cite{foot2}, which is equal to the total chirality of the Weyl points enclosed. In Fig.~\ref{pump}(c), $\phi_{y,0}=\frac{\pi}{2}$, $\phi_{z,0}=\arccos\left(\frac{\pi}{V}\right)$, and $r=\frac{\pi}{10}$ so that the adiabatic loop encloses two Weyl points at quasienergy $\pi$ at $(k,\phi_y,\phi_z)=\left(\pm\frac{\pi}{2},\frac{\pi}{2},\arccos\left(\frac{\pi}{V}\right)\right)$, and $\frac{\Delta\langle X \rangle}{2a}=2$, which is equal to the total chirality of the Weyl points enclosed multiplied by minus one. An interesting result is observed in Fig.~\ref{pump}(d). Under $\phi_{y,0}=\frac{\pi}{2}$, $\phi_{z,0}=\frac{\pi}{4}+\frac{\arccos\left(\frac{\pi}{V}\right)}{2}$, and $r=\frac{3\pi}{10}-\frac{\arccos\left(\frac{\pi}{V}\right)}{2}$, the adiabatic loop encloses two Weyl points at quasienergy $0$ at $(k,\phi_y,\phi_z)=\left(\pm\frac{\pi}{2},\frac{\pi}{2},\frac{\pi}{2}\right)$ and two other Weyl points at quasienergy $\pi$ at $(k,\phi_y,\phi_z)=\left(\pm\frac{\pi}{2},\frac{\pi}{2},\arccos\left(\frac{\pi}{V}\right)\right)$. Although all the Weyl points enclosed have the same chirality $-1$, $\frac{\Delta\langle X \rangle}{2a}=0$ rather than $4$. This suggests that the adiabatic pumping scheme we considered responds to the chirality of the Weyl points at quasienergy $0$ and quasienergy $\pi$ in the opposite manner. These numerical results agree with the two observations earlier based on analytical derivation.

\subsection{$\pi$ Berry phase around line nodes}

\begin{figure}
\begin{center}
\includegraphics[scale=0.5]{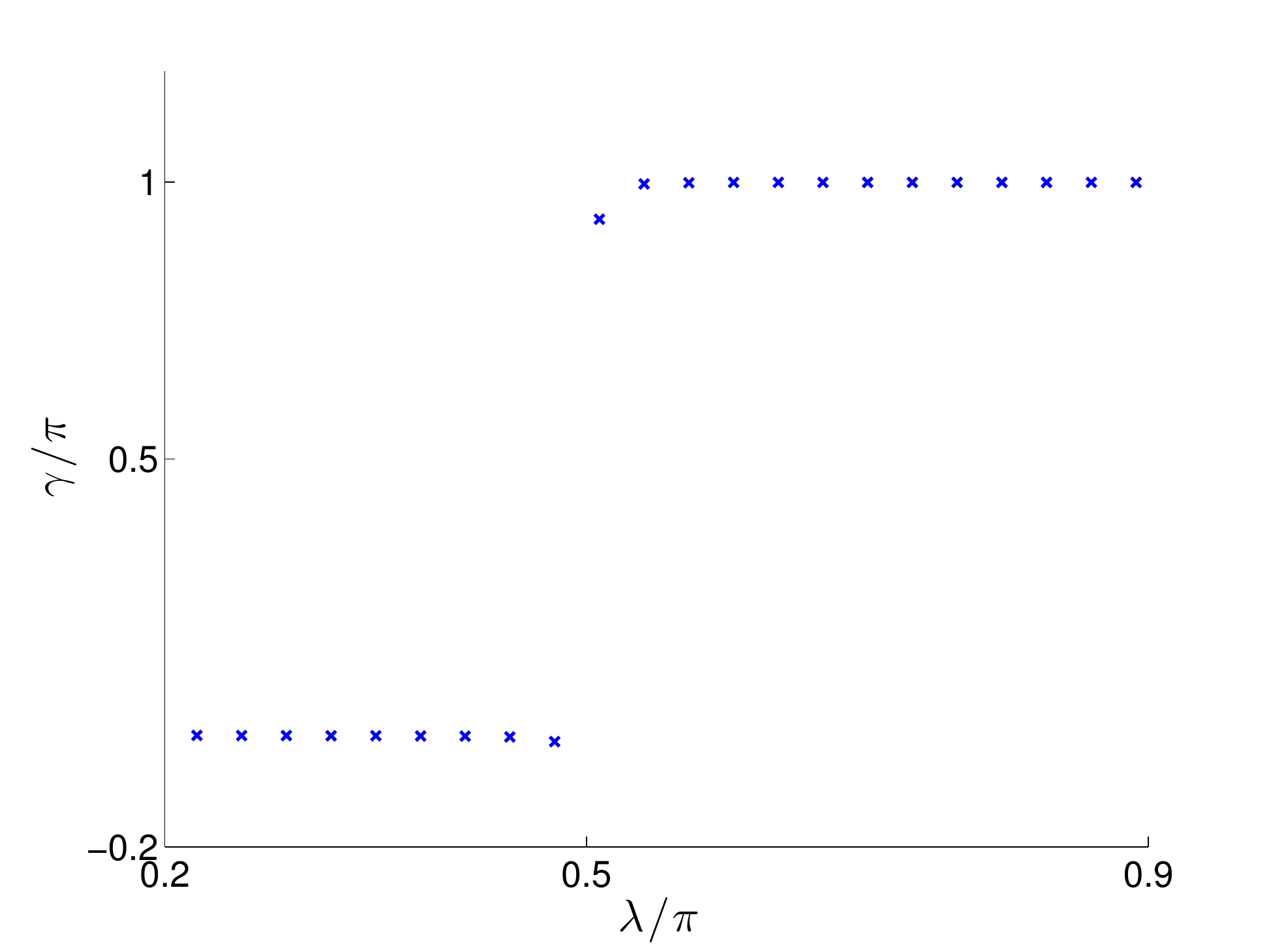}
\end{center}
\caption{Berry phase acquired by a Floquet eigenstate~(at a fixed $k=\frac{\pi}{2}$) adiabatically driven along a closed loop in the artificial dimensions as a function of $\lambda$ at $J=1$ and $V=2$.}
\label{piberry}
\end{figure}

As suggested in Sec.~\ref{II} B, the line nodes appearing in the ODKHM can be characterized by a topological winding number, which is closely related to the Berry phase around a loop enclosing a line node. This $\pi$ Berry phase can also be manifested via the following adiabatic pumping scheme. First, we prepare a Floquet eigenstate at a fixed $k=\frac{\pi}{2}$ as an initial state. Then, we drive it adiabatically along a closed loop in the artificial dimensions parametrized by $\theta$ according to $\phi_y=\frac{\pi}{2}+\frac{\pi}{5}\cos\left[\theta(t)\right]$ and $\phi_z=\frac{\pi}{2}+\frac{\pi}{5}\sin\left[\theta(t)\right]$. Fig.~\ref{piberry} shows the Berry phase at $J=1$ and $V=2$ as $\lambda$ is varied from $0.5$ to $2.9$. It shows that when $0.5<\lambda<\frac{\pi}{2}$, the Berry phase is $0$ as the loop does not enclose any line node at these values of $\lambda$. When $\lambda>\frac{\pi}{2}$ and therefore line nodes start to appear, the Berry phase jumps to $\pi$.

We note also that the $\pi$ Berry phase obtained above when $\lambda>\frac{\pi}{2}$ is contributed only due to the singularity of the line node, i.e., the point where the two quasienergies become degenerate, and hence due to its topology. This is because the contribution from the integral of the Berry curvature over the area of the loop, excluding the singularity due to the line node, is $0$. To show this, we start by rewriting Eq.~(\ref{mod}) as follows,

\begin{equation}
\mathcal{U}\left(k=\frac{\pi}{2},k_y,k_z\right)=e^{2\mathrm{i} \lambda \sin(k_y)\sigma_y}e^{\mathrm{i}V\sin(k_z)\sigma_z}\;,
\label{med}
\end{equation}

\n where $k_y=\frac{\pi}{2}+\phi_y$ and $k_z=\frac{\pi}{2}+\phi_z$. We then apply a unitary transformation to Eq.~(\ref{med}) to obtain,

\begin{equation}
\mathcal{U'}\left(k=\frac{\pi}{2},k_y,k_z\right)=e^{\mathrm{i} \lambda \sin(k_y)\sigma_y}e^{\mathrm{i}V\sin(k_z)\sigma_z}e^{\mathrm{i} \lambda \sin(k_y)\sigma_y}\;.
\label{med2}
\end{equation}

\n Since $\mathcal{U}\left(k=\frac{\pi}{2},k_y,k_z\right)$ and $\mathcal{U'}\left(k=\frac{\pi}{2},k_y,k_z\right)$ are related by a unitary transformation, they share the same eigenvalues, while the eigenvectors are related by

\begin{equation*}
\left|u\left(k=\frac{\pi}{2},k_y,k_z\right)\right\rangle =e^{\mathrm{i}\lambda \sin(k_y)\sigma_y}\left|u'\left(k=\frac{\pi}{2},k_y,k_z\right)\right\rangle \;.
\end{equation*}

\n Furthermore, Eq.~(\ref{med2}) implies that $\mathcal{U'}\left(k=\frac{\pi}{2},k_y,k_z\right)=\mathcal{U'}^\dagger\left(k=\frac{\pi}{2},-k_y,-k_z\right)$. It therefore follows that if $\left|u'^+\left(k=\frac{\pi}{2},k_y,k_z\right)\right\rangle$ and $\left|u'^-\left(k=\frac{\pi}{2},k_y,k_z\right)\right\rangle$ are eigenvectors to $\mathcal{U'}\left(k=\frac{\pi}{2},k_y,k_z\right)$ that correspond to quasienergy $+\varepsilon\left(k=\frac{\pi}{2},k_y,k_z\right)$ and $-\varepsilon\left(k=\frac{\pi}{2},k_y,k_z\right)$ respectively,

\begin{equation*}
\left|u'^\pm\left(k=\frac{\pi}{2},k_y,k_z\right)\right\rangle = \left|u'^\mp\left(k=\frac{\pi}{2},-k_y,-k_z\right)\right\rangle \;.
\end{equation*}

\n In terms of the eigenvectors of $\mathcal{U}\left(k=\frac{\pi}{2},k_y,k_z\right)$,

\begin{equation*}
\left|u^\pm\left(k=\frac{\pi}{2},k_y,k_z\right)\right\rangle=e^{2\mathrm{i}\lambda \sin(k_y)\sigma_y}\left|u^\mp\left(k=\frac{\pi}{2},-k_y,-k_z\right)\right\rangle \;.
\end{equation*}

\n Therefore, the Berry curvature at $\left(k,\phi_y,\phi_z\right)=\left(\frac{\pi}{2}, \frac{\pi}{2}+k_y,\frac{\pi}{2}+k_z\right)$ can be shown to be

\begin{equation*}
\mathcal{B}_\pm(k_y,k_z)=-\mathcal{B}_\pm(-k_y,-k_z) \;,
\end{equation*}

\n which is odd in $k_y$ and $k_z$, and thus its integral over the area, excluding the singularity due to the line node, enclosed by the loop we defined earlier is $0$.

\section{Concluding remarks}\label{V}

In this study, we have shown that even by focusing only on the two-band case, the Floquet states of the ODKHM are able to host as many Weyl points and line nodes as possible. By taking OBC along the physical dimension of the system, dispersionless edge states in a form of a line connecting two Weyl points with opposite chiralities are observed, which is analogue to the Fermi arc in static systems. The emergence of these edge states can be connected with a type of $1$D topological invariant, such as that used to characterize the topology of the SSH model. Although the edge states associated with the line nodes cannot be observed when the system is open along its physical dimension, they become visible if instead we open the system along its artificial dimension.

By preparing a Wannier state centered around $n=0$ as the initial state, and driving it adiabatically along a closed loop in its artificial dimensions, which can be achieved by simply changing the experimental parameters $\phi_y$ and $\phi_z$, the change in position expectation value $\Delta \langle X \rangle$ of the wave packet is quantized by the total chirality of the Weyl points enclosed. We further observed that under the same initial state and adiabatic pumping scheme, the chirality of the Weyl points at quasienergy $0$ and quasienergy $\pi$ contributes to $\Delta \langle X \rangle$ in an opposite manner. Lastly, the $\pi$ Berry phase of the line nodes is verified by adiabatically driving a Floquet eigenstate along a closed loop in the artificial dimensions.

This paper should therefore add new knowledge on the studies of the topological semimetal phases in Floquet systems. Following this study, there are many possible future directions which deserve further explorations. For example, given that the two-band ODKHM has a very rich Floquet band structure, it is expected that considering the ODKHM with more than two bands may give rise to more interesting novel phenomena to observe. Furthermore, we might also consider a continuous driven version of this model by replacing the kicking strength in Eq.~(\ref{fund}) with a harmonic driving amplitude~\cite{Longwen}, which is more feasible to realize experimentally. In addition, it will be interesting to further explore the difference between Weyl points at quasienergy $0$ and quasienergy $\pi$, for example in terms of their edge states behaviour or chiral anomaly~\cite{Hosur, Burkov2}. Lastly, since we can practically control the number of edge states in the two-band ODKHM, it will be a good platform to study interaction effects and edge state transfer~\cite{Doublon} in Floquet systems.

\vspace{1cm}

\n {\bf Acknowledgements:} We thank Hailong Wang and Dario Poletti for helpful discussions. This work is supported by Singapore Ministry of Education Academic Research Fund Tier I~(WBS No. R-144-000-353-112).

\appendix

\vspace{1cm} \begin{center} {\bf Appendix} \end{center}

This Appendix has four sections. In Appendix A, we present detailed derivations of Eq.~(4) in the main text. This is followed by Appendix B, where Lemma $1$ is proved. In Appendix C, we derive the approximate Floquet operator around different types of band touching points. In Appendix D, we establish the relation between the quantized adiabatic pumping of Wannier states and the chirality of the Weyl nodes. Whenever possible, we use the same notation as in the main text.

\section{Derivation of Eq.~(\ref{mod})}
\label{derive}

Define $|a_n \rangle = |2 n \rangle$ and $|b_n \rangle = |2n-1 \rangle$. In terms of $|a_n \rangle$ and $|b_n \rangle$, Eq.~(\ref{tri}) can be written as,

\begin{equation}
U=e^{-\mathrm{i} H_\mathrm{OFF}}e^{-\mathrm{i} H_\mathrm{ON}}\;,
\end{equation}

\n where,

\begin{eqnarray}
	\label{off} H_\mathrm{OFF} &=& \sum_{n} \left\lbrace \left[J+\lambda \cos(\phi_y)\right]|b_{n-1} \rangle \langle a_n |+ \left[J-\lambda \cos(\phi_y)\right]|a_n \rangle \langle b_n | + h.c\right\rbrace ,  \\
	H_\mathrm{ON} &=&  \sum_{n} \left[ V \cos(\phi_z) |a_n\rangle \langle a_n | - V \cos(\phi_z) |b_n\rangle \langle b_n |\right] . \label{on}
\end{eqnarray}

\n Next, introduce Fourier transformed quantities,

\begin{eqnarray}
	|a_n \rangle &=& \frac{1}{\sqrt{\frac{N}{2}}} \sum_k e^{-\mathrm{i} k (2n) c} |a_k\rangle , \\
	|b_n \rangle &=& \frac{1}{\sqrt{\frac{N}{2}}} \sum_k e^{-\mathrm{i} k (2n-1) c} |b_k\rangle ,
	\end{eqnarray}
	
\n	where $c$ is the lattice constant, which can be taken to be $1$ for simplicity. Eq.~(\ref{off}) and~(\ref{on}) become,

\begin{eqnarray}
	H_\mathrm{OFF} &=& \sum_k \left[ (J+\lambda \cos(\phi_y))e^{-\mathrm{i} k}|b_k \rangle \langle a_k |+ (J-\lambda \cos(\phi_y))e^{-\mathrm{i} k}|a_k \rangle \langle b_k | + h.c\right] \nonumber \\
	& \hat{=} & \sum_k \left[\begin{array}{cc} 0 & 2 J \cos(k)-2\mathrm{i}\lambda \sin(k) \cos(\phi_y) \\ 2 J \cos(k)+2\mathrm{i}\lambda \sin(k) \cos(\phi_y) & 0 \\ \end{array}\right] \otimes |k \rangle \langle k |\nonumber\\
	&=& \sum_k \left(2 J \cos(k) \sigma_x+2\lambda \sin(k)\cos(\phi_y) \sigma_y \right)\otimes |k\rangle \langle k | \nonumber \\
	&=& \sum_k h_\mathrm{OFF}(k) \otimes |k \rangle \langle k | , \\
H_\mathrm{ON}&=& \sum_{k} V \cos(\phi_z) |a_k\rangle \langle a_k | - V \cos(\phi_z) |b_k\rangle \langle b_k | \nonumber \\
&\hat{=}& \sum_k \left[\begin{array}{cc} V\cos(\phi_z) & 0 \\ 0 & -V \cos(\phi_z) \\ \end{array}\right] \otimes |k\rangle \langle k | \nonumber \\
&=& \sum_k V \cos(\phi_z) \sigma_z\otimes |k \rangle \langle k | \nonumber \\
&=& \sum_k h_\mathrm{ON}(k) \otimes |k\rangle \langle k |.
\end{eqnarray}

\n $k$ is therefore a conserved quantity, and Eq.~(\ref{tri}) becomes,

\begin{eqnarray}
U &=& \sum_k \left[e^{-\mathrm{i}h_\mathrm{OFF}(k)}e^{-\mathrm{i}h_\mathrm{ON}(k)}\right]\otimes |k\rangle \langle k | \nonumber \\
&=& \sum_k \mathcal{U}(k,\phi_y,\phi_z) \otimes |k \rangle \langle k |,
\end{eqnarray}

\n where $\mathcal{U}(k,\phi_y,\phi_z)$ is given by Eq.~(\ref{mod}).

\section{Proof of Lemma 1}
\label{proof}
	We start by using Euler's Formula,
	\begin{eqnarray*}
	e^{\mathrm{i} a \vec{n}\cdot \sigma} e^{\mathrm{i} b \vec{m}\cdot \sigma} &=& \left[\cos(a)+\mathrm{i} \sin(a) \vec{n}\cdot \sigma \right] \left[\cos(b)+\mathrm{i}\sin(b) \vec{m}\cdot \sigma \right] \\
	&=& \cos(a)\cos(b)-\sin(a)\sin(b)\vec{n}\cdot\vec{m} +\mathrm{i} \left[ \sin(a)\cos(b) \vec{n}\cdot \sigma\right. \\
	&& \left.+\sin(b)\cos(a) \vec{m}\cdot \sigma +\sin(a)\sin(b)\vec{m}\times\vec{n}\cdot \sigma \right] \\
	&=& e^{\mathrm{i}\theta \vec{r}\cdot \sigma},
	\end{eqnarray*}
	
	\n where $r$ is some unit vector and $\theta = \arccos\left[\cos(a)\cos(b)-\sin(a)\sin(b)\vec{n}\cdot\vec{m}\right]$. Since $\vec{n}$ and $\vec{m}$ are perpendicular, $\theta = \arccos\left[\cos(a)\cos(b)\right]$. Band touching points correspond to $\theta$ being an integer multiple of $\pi$. Since cosine function is bounded in $\left[ -1, 1\right]$, this is only possible if both $\cos(a)$ and $\cos(b)$ are $\pm 1$. This proves Lemma \ref{lem}.

\section{Expansion of the Floquet operator around band touching points}
\begin{center}
\item \subsection{Around band touching points of type $1$ (Weyl points)}
\end{center}

As explained in the main text, one of the band touching points occurs at $(k,\phi_y,\phi_z)=\left(\frac{\pi}{2},\frac{\pi}{2},\phi_1\right)$, where $\phi_1=\arccos\left(\frac{l_2\pi}{V}\right)$. Let $k_x=k-\frac{\pi}{2}$, $k_y=\phi_y-\frac{\pi}{2}$, and $k_z=\phi_z-\phi_1$. Eq.~(\ref{mod}) then becomes,

\begin{equation}
\mathcal{U}(k_x,k_y,k_z) = e^{\mathrm{i} (2J k_x \sigma_x +2\lambda k_y\sigma_y)} e^{\left[-\mathrm{i} l_2 \pi +\mathrm{i}V\sin(\phi_1)k_z\right]\sigma_z} \;.
\label{c1}
\end{equation}

\n The second exponential can be written as $e^{-\mathrm{i} l_2 \pi \sigma_z}e^{\mathrm{i}V\sin(\phi_1)k_z\sigma_z}$ since $\sigma_z$ commutes with itself. Furthermore, according to Euler's formula, $e^{-\mathrm{i} l_2 \pi \sigma_z}=\cos(l_2\pi)+\mathrm{i}\sin(l_2\pi)\sigma_z=\pm 1=e^{-\mathrm{i}l_2\pi}$, where the sign depends on whether $l_2$ is an even or odd integer. In both cases, the resulting exponential commutes with all Pauli matrices, and Eq.~(\ref{c1}) becomes,

\begin{eqnarray*}
\mathcal{U}(k_x,k_y,k_z) &=& e^{-\mathrm{i}l_2 \pi}e^{\mathrm{i} (2J k_x \sigma_x +2\lambda k_y\sigma_y)} e^{\mathrm{i}V\sin(\phi_1)k_z\sigma_z} \\
&\approx & e^{-\mathrm{i}l_2 \pi}e^{\mathrm{i} \left[2J k_x \sigma_x +2\lambda k_y\sigma_y+V\sin(\phi_1)k_z\sigma_z\right]} \\
&=& e^{-\mathrm{i} \left\lbrace l_2\pi- \left[2J k_x \sigma_x +2\lambda k_y\sigma_y+V\sin(\phi_1)k_z\sigma_z\right]\right\rbrace} \\
\end{eqnarray*}

\begin{center}
\item \subsection{Around line nodes of type $2$ or $3$}
\end{center}

Since the derivation of the effective Hamiltonian near a line node of type $2$ and $3$ is similar, we will present the derivation of Eq.~(\ref{lineff}) only in this section. Let $k_x=k-k_0'$, $k_y=\phi_y-\frac{\pi}{2}$, and $k_z=\phi_z-\frac{\pi}{2}$, where $k_0'=\arccos\left(\frac{\pi}{2J}\right)$. Eq.~(\ref{mod}) then becomes,

\begin{equation}
\mathcal{U}(k_x,k_y,k_z) = e^{\mathrm{i} \left\lbrace\left[-\pi + 2J k_x\sin(k_0')\right] \sigma_x +2\lambda k_y\sin(k_0')\sigma_y\right\rbrace} e^{\mathrm{i}V\sin(\phi_1)k_z\sigma_z} \;.
\label{c2}
\end{equation}

\n By using Euler's formula, Eq.~(\ref{c2}) can be written as,

\begin{eqnarray}
\mathcal{U}(k_x,k_y,k_z) &=& \cos(\theta)\cos(V k_z)+\mathrm{i} \left[ \cos(\theta)\sin(Vk_z)\sigma_z\right. \nonumber\\
&& +\frac{\sin(\theta)}{\theta}\left\lbrace -2\lambda k_y\sin(k_0') \sin(Vk_z)+ \left[2Jk_x\sin(k_0')-\pi\right]\cos(Vk_z)\right\rbrace\sigma_x\nonumber \\
&& +\frac{\sin(\theta)}{\theta}\left\lbrace 2\lambda k_y\sin(k_0') \cos(Vk_z)+\left[2Jk_x\sin(k_0')-\pi\right]\sin(Vk_z)\right\rbrace \sigma_y\left.\right] \nonumber \\
&=& e^{\mathrm{i}\frac{\theta'}{\sin(\theta')}\vec{r}\cdot \sigma} \nonumber \\
&=& e^{-\mathrm{i}h_\mathrm{eff}} ,
\end{eqnarray}

\n where $\theta=\sqrt{\left[\pi-2Jk_x\sin(k_0')\right]^2+4\lambda^2k_y^2\sin^2(k_0')}$ and $\theta' = \arccos\left[\cos(\theta)\cos(Vk_z)\right]$. Up to first order in $k_x$, $k_y$, and $k_z$, $\theta' \approx \pi-\sqrt{V^2k_z^2+4J^2k_x^2\sin^2(k_0')}=\pi-\Phi(k_x,k_z)$, $\sin(\theta')\approx -\Phi(k_x,k_z)$, $\sin(\theta)\approx-2Jk_x\sin(k_0')$, and $\frac{\theta'}{\theta}\approx 1-\frac{\Phi(k_x,k_z)-2Jk_x\sin(k_0')}{\pi}$. It therefore follows that,

\begin{eqnarray}
h_\mathrm{eff} &=& -\frac{\theta'}{\sin(\theta')}\vec{r}\cdot \sigma \nonumber \\
&=& \left[\pi-\Phi(k_x,k_z)\right]\left[-\frac{Vk_z}{\Phi(k_x,k_z)}\sigma_z+\frac{2Jk_x\sin(k_0')}{\Phi(k_x,k_z)}\sigma_x\right] \nonumber \\
&& +\frac{2Jk_x\sin(k_0')}{\Phi(k_x,k_z)}\left[-2\lambda k_y\sin(k_0')+\pi Vk_z\right]\sigma_y \nonumber \\
&=& \left[\pi-\Phi(k_x,k_z)\right]\hat{n}\cdot\sigma +\frac{2Jk_x\sin(k_0')}{\Phi(k_x,k_z)}[-2\lambda k_y\sin(k_0')+\pi Vk_z]\sigma_y ,
\end{eqnarray}

\n as given in the main text. The derivation of Eq.~(\ref{lineff2}) follows exactly the same step as above with $\sigma_y\rightarrow \sigma_x$, $\sigma_x\rightarrow \sigma_y$, $\lambda k_y\sin(k_0')\rightarrow Jk_x$, and $Jk_x\sin(k_0')\rightarrow \lambda k_y \sin(\phi_0)$.

\vspace{5mm}

\section{Analytical derivation of the quantized particle transport around Weyl points}
\label{quantized}

The purpose of this section is to show how chirality emerges when an adiabatic pumping scheme elucidated in Sec.~\ref{IV} A around a Weyl point is performed. A typical effective Hamiltonian around a Weyl point in the ODKHM takes the following form,

\begin{equation}
H_\mathrm{eff}=l\pi - v_x k_x\sigma_x -v_y k_y\sigma_y -v_z k_z \sigma_z\;.
\label{Weyl}
\end{equation}

\n Since quasienergy is only defined mod $2\pi$, odd $l$ corresponds to a Weyl point at quasienergy $\pi$ while even $l$ corresponds to a Weyl point at quasienergy $0$. By parametrizing $k_y=r\cos(\theta)$ and $k_z=r\sin(\theta)$, the lower-band eigenstates as a function of $k,\theta\in [-\pi,\pi)$ associated with the above effective Hamiltonian at quasienergy $\pi$ and $0$ respectively can be explicitly calculated in the $\sigma_y$ representation as follow,

\begin{eqnarray}
|\pi (k_x,\theta) \rangle &\hat{=}& \frac{1}{\sqrt{2r_0^2+2v_x^2 k_x^2+2 v_x k_x \sqrt{r_0^2+v_x^2 k_x^2}}}\left( \begin{array}{c} r_0 e^{-\mathrm{i} \phi} \\ - v_x k_x - \sqrt{r_0^2+v_x^2 k_x^2}\end{array}\right) , \\
|0 (k_x,\theta) \rangle &\hat{=}& \frac{1}{\sqrt{2r_0^2+2v_x^2 k_x^2- 2v_x k_x \sqrt{r_0^2+v_x^2 k_x^2}}}\left( \begin{array}{c} r_0 e^{-\mathrm{i} \phi} \\ - v_x k_x + \sqrt{r_0^2+v_x^2 k_x^2}\end{array}\right) ,
\end{eqnarray}

\n where $r_0 = r\sqrt{v_y^2 \cos^2(\theta)+ v_z^2 \sin^2(\theta)}$ and $\tan(\phi) = \frac{v_z}{v_y}\tan(\theta)$. By preparing an initial state in a form of Eq.~(\ref{is}) with $k_x$, $k_y$, and $k_z$ replace $k$, $\phi_y$, and $\phi_z$ respectively, both the population weight of the initial state and the quasienergy spectrum of Eq.~(\ref{Weyl}) are invariant under reflection about $k_x=0$. Ref.~\cite{Hailong} then suggests that the shift in position expectation value after one adiabatic cycle is given by,

\begin{eqnarray}
\Delta \langle X \rangle_{\pi(0)} &=& \frac{a \mathrm{i}}{2\pi} \int_{-\frac{\pi}{a}}^\frac{\pi}{a} dk_x \int_0^{2\pi} d\theta  \left[ \langle \partial_{k_x}\pi(0) | \partial_\theta \pi(0) \rangle - c.c. \right] \nonumber \\
&=& \frac{a}{2\pi} \int_{-\frac{\pi}{a}}^\frac{\pi}{a} dk_x \int_0^{2\pi} d\theta \frac{d\phi}{d\theta} \frac{d}{dk_x}\left[\frac{r_0^2}{2r_0^2+2v_x^2 k_x^2+(-) 2v_x k_x \sqrt{r_0^2+v_x^2 k_x^2}}\right]\nonumber \\
&=& \frac{a}{2\pi} \int_0^{2\pi} d\theta \frac{d\phi}{d\theta} \left\lbrace \int_{-\frac{\pi}{a}}^\frac{\pi}{a} dk_x \frac{d}{dk_x}\left[\frac{r_0^2}{2r_0^2+2v_x^2 k_x^2+(-) 2v_x k_x \sqrt{r_0^2+v_x^2 k_x^2}}\right] \right\rbrace. \label{save}
\end{eqnarray}

Let's first focus on the term inside the square bracket. Since we can take $r$ as small as possible, and hence $r_0$ as small as possible, we can assume $\frac{\pi}{a}\gg r_0$, so that upon performing the integration,

\begin{eqnarray*}
\int_{-\frac{\pi}{a}}^\frac{\pi}{a} dk_x \frac{d}{dk_x}\left[\frac{r_0^2}{2r_0^2+2v_x^2 k_x^2+(-) 2v_x k_x \sqrt{r_0^2+v_x^2 k_x^2}}\right] &=& \frac{r_0^2}{2r_0^2+2v_x^2 \frac{\pi^2}{a^2}+(-) 2v_x \frac{\pi}{a} \sqrt{r_0^2+v_x^2 \frac{\pi^2}{a}}}\\
&&-\frac{r_0^2}{2r_0^2+2v_x^2 \frac{\pi^2}{a^2}-(+) 2v_x \frac{\pi}{a} \sqrt{r_0^2+v_x^2 \frac{\pi^2}{a}}} \\
&\approx & -(+) \frac{4 r_0^2 \frac{v_x\pi}{a} \sqrt{r_0^2+v_x^2 \frac{\pi^2}{a}}}{\left(2r_0^2+2\frac{v_x^2\pi^2}{a}\right)^2-4\frac{v_x^2\pi^2}{a}\left(r_0^2+v_x^2 \frac{\pi^2}{a}\right)} \\
&\approx & -(+) \mathrm{sgn}(v_x)
\end{eqnarray*}

\n Therefore, Eq.~(\ref{save}) becomes,

\begin{eqnarray}
\Delta \langle X \rangle_{\pi(0)} &=& -(+) \frac{a}{2\pi} \int_0^{2\pi} d\theta \frac{d\phi}{d\theta} \mathrm{sgn}(v_x) \nonumber \\
&=& -(+) \frac{a}{2\pi} \mathrm{sgn}(v_x) \left[\phi(\theta =2\pi)-\phi (\theta =0)\right] \nonumber \\
&=& -(+) a \mathrm{sgn}(v_x) \mathrm{sgn}(v_z)\mathrm{sgn}(v_y) \nonumber \\
&=& -(+) a\mathrm{sgn}(v_xv_yv_z) \nonumber \\
&=& +(-) a \chi .
\label{chiral}
\end{eqnarray}

We have thus showed that the change in position expectation value after one adiabatic cycle is quantized by the lattice constant times the chirality of the Weyl point enclosed. Although in the derivation above we assumed that the radius $r$ of the loop in the artificial dimensions is very small, the above result holds for arbitrary $r$ or even arbitrary shape of the loop enclosing the Weyl point, provided there are no other singularities inside the loop we constructed. This is because chirality can also be interpreted as the total Berry flux passing through a given closed surface. In analogy to Gauss Law in electrostatics, the total flux should not depend on the shape of the closed surface. Therefore, as long as the initial state fully populates all quasimomenta $k$~(so that $k$ and $\theta$ form a closed surface) and has a reflection symmetry about $k_x=0$, the above result is expected to hold.

\label{exp}


\begin{thebibliography}{99}
\bibitem{foot1} Since the quasi-energy spectrum is only defined mod $2\pi$, it is not clear which band is lower or higher. Here, by lower band I'm referring to the values of quasi-energy that belong to the same equivalence class as quasi-energy smaller than $0$.
\bibitem{foot2} The factor $2$ in the denominator of $\frac{\Delta \langle X \rangle}{2a}$ comes from the fact that Eq.~(\ref{tri}) is periodic by two lattice sites. The effective lattice constant of the system appearing in Eq.~(\ref{chiral}) is therefore $2a$.
\end{thebibliography}
\end{document}